\documentclass[nojss]{jss}

\usepackage{listings}
\usepackage{url}
\usepackage{amsmath}
\usepackage{amssymb}
\usepackage{bbm}
\usepackage{color}
\usepackage{float}
\definecolor{grey}{gray}{0.9}
\definecolor{Orange}{rgb}{1,0.5,0}

\makeatletter
\g@addto@macro{\UrlBreaks}{\UrlOrds}
\makeatother

\DeclareMathOperator*{\argmax}{arg\,max}

\author{Jules Kruijswijk\\Tilburg University \And
	Robin van Emden\\Jheronimus Academy \\ of Data Science \And
	Petri Parvinen\\Aalto University \And
	Maurits Kaptein\\Tilburg University}
\title{\pkg{StreamingBandit}; Experimenting with Bandit Policies}

\Plainauthor{Jules Kruijswijk, Robin van Emden, Petri Parvinen, Maurits Kaptein} 
\Plaintitle{StreamingBandit; Experimenting with Bandit Policies} 
\Shorttitle{\pkg{StreamingBandit}} 

\Abstract{A large number of statistical decision problems in the social sciences and beyond can be framed as a (contextual) multi-armed bandit problem. However, it is notoriously hard to develop and evaluate policies that tackle these types of problem, and to use such policies in applied studies. To address this issue, this paper introduces \pkg{StreamingBandit}, a \proglang{Python} web application for developing and testing bandit policies in field studies. \pkg{StreamingBandit} can sequentially select treatments using (online) policies in real time. Once \pkg{StreamingBandit} is implemented in an applied context, different policies can be tested, altered, nested, and compared. \pkg{StreamingBandit} makes it easy to apply a multitude of bandit policies for sequential allocation in field experiments, and allows for the quick development and re-use of novel policies. In this article, we detail the implementation logic of \pkg{StreamingBandit} and provide several examples of its use. 
}

\Plainkeywords{Sequential decision-making, Multi-armed bandit, data streams, sequential experimentation, Python}

\Address{
  Jules M.A. Kruijswijk\\
  Tilburg University\\
  Statistics and Research Methods\\
  Tilburg, the Netherlands \\
  E-mail: \email{j.m.a.kruijswijk@uvt.nl}\\
  \linebreak
  Robin van Emden\\
  JADS Den Bosch\\
  Data Science\\
  Den Bosch, the Netherlands\\
  E-mail: \email{r.a.vanemden@uvt.nl}\\
  \linebreak
  Petri Parvinen\\
  Aalto University\\
  Industrial Engineering \& Management\\
  Helsinki, Finland\\
  E-mail: \email{petri.parvinen@aalto.fi}\\
  \linebreak
  Maurits C. Kaptein\\
  Tilburg University \& JADS\\
  Statistics and Research Methods\\
  Tilburg, the Netherlands\\
  E-mail: \email{m.c.kaptein@uvt.nl}\\
}

\begin{document}

\section{Introduction}

In the canonical multi-armed bandit (MAB) problem a gambler faces a number of slot machines, each with a potentially different payoff. It is the gambler's goal to make as much profit (or, in the case of gambling, as little loss) as possible by sequentially choosing which machine to play, learning from the observations as she goes along \citep{Whittle1980, berry1985bandit}. The gambler faces a trade-off between exploration and exploitation: on the one hand she wishes to play the machine that was successful in earlier attempts as often as possible (exploitation), but on the other hand she wishes to find the machine with the highest payoff through experimentation (exploration) \citep{Macready1998}. The MAB problem, and its generalization, the contextual MAB (or CMAB) problem---in which before selecting a machine the gambler observes the state of the world that could be related to the optimal choice of machine at that point in time---provides a flexible formalization for studying sequential treatment-allocation procedures in the social sciences and beyond \citep{Dudik2011a, Li2010a, Agrawal2014}.

A multitude of policies addressing (contextual) decision problems have been conceived and evaluated \citep[see,  e.g.,][]{berry1985bandit, Chapelle2011, Dudik2011a}. Indeed, the randomized controlled trial (RCT, or $\epsilon$-first in the literature on sequential decision-making \citep{Chapelle2011}) is in itself a specific policy devised to address the exploration-exploitation trade-off in which an exploration phase, the trial itself, is followed by exploitation. Other policies range from simple heuristics such as ``play the winner"  \citep{Lachin1988,  Villar2015} to asymptotically optimal policies such as Upper Confidence Bound (UCB) methods \citep{Auer2010, Garivier2011, Audibert2009}, and Bayesian methods such as Thompson sampling \citep{thompson1933likelihood, Chapelle2011, Agrawal2011}. It is difficult to assess which of these policies performs best in distinct applied problems, however, due to the omission of the counterfactuals in the (field) evaluations of a policy \citep{Li2010a}: one does not know what the outcome would have been had another choice been made anywhere along the sequence of decisions. Hence the data resulting from an evaluation can often not be used to evaluate alternative policies. To evaluate a range of possible policies one has to resort to either simulation methods---which often lack external validity due to the large number of assumptions encoded in the simulation---or to recent offline evaluation methods \citep{Li2010a, agarwal2016making}. Offline methods provide the opportunity to obtain unbiased estimates of the performance of different policies on historical data, but these approaches are only practically feasible when the number of choice alternatives is relatively low and/or the number of sequential choices is large. Furthermore, the assumptions that justify these methods---such as stationarity and a non-zero probability for each possible treatment at each interaction \citep{Li2010a}---are rarely fully justified in practice.

Despite these difficulties, effective (contextual) decision policies are potentially of great use in many areas. To unleash this potential researchers need to be able quickly to implement and evaluate distinct bandit policies in the field. This can be achieved by allowing substantive researchers easily to test different sequential allocation schemes. If easy-to-use software were available for evaluating and disseminating novel policies, such policies---which are actively being developed \citep[e.g.,][]{Kaptein, Osband2015, bastani2015online}---would be within reach of a broader research community. It is to this end that we developed \pkg{StreamingBandit}: an open-source RESTful Web application that allows researchers to formalize their sequential-allocation procedure as a CMAB problem and, by virtue of this formalization, easily to experiment with different policies.  

In the remainder of this section we first engage in a high-level discussion of the basic usage of \pkg{StreamingBandit}, discuss related approaches, and provide an overview of the application and its installation. In section \ref{sec:usage} we describe the application in more detail, and demonstrate the setup and evaluation of a single policy. Here we also discuss the use of \pkg{StreamingBandit} for offline policy evaluation and we offer a number of performance measures. In section \ref{sec:examples} we introduce a number of currently implemented ``default" policies and discuss methods of combining multiple policies. We detail two practical applications of \pkg{StreamingBandit} in section \ref{sec:practice}, and finally in section \ref{sec:discussion} we briefly discuss future work directions. 

\subsection{Basic Usage}

The basic setting we consider is the following. Consider an experimenter who interacts with the environment. At each interaction $t$:
\begin{enumerate}
\item the experimenter observes a context $x_t$,
\item subsequently, the experimenter chooses an action $a_t$,
\item and finally a reward $r_t$ is observed.
\end{enumerate}
The main aim of the experimenter is to maximize the cumulative reward $\sum_{t=1}^N r_t$ where $N$ denotes the total number of interactions. To do so, the experimenter applies a policy $\pi$ which is some function that takes the context $x_t$ and the historical interactions, and returns an action. For convenience we denote all historical interactions using $\mathcal{D}_{(t-1)}$ and thus we have $\pi(x_t, \mathcal{D}_{t-1}) \rightarrow a_t$. 

This sequential decision-making scheme is encountered in many real-life situations:
\begin{itemize}
\item Personalized healthcare: A physician meets with patients sequentially. For each patient, she observes a number of background characteristics (gender, age, current condition) constituting the context. Subsequently, her action is to choose a treatment such that the reward---measured in terms of the general health of the patient---is maximized.
\item Online advertising: In online advertising a firm selecting an ad observes the context consisting of a description of the current user visiting a specific webpage. The action is to choose an advertisement from of a set of possible advertisements (possibly dependent on the context), and the rewards constitute the clicks on the ad.
\item Product-recommendation systems: The context denotes all that is known about the user at a certain point in time. The action is choosing one of a set of products, and the reward consists of the revenue generated at each interaction.
\item Social-science experiments: Many social-science experiments constitute a special case of contextual decision-making: participants are recruited sequentially during the experiment. The context consists of all that is known about the participant, and sequentially the action is to assign a participant to a specific experimental condition (possibly dependent on the context in cases of stratified sampling, for example). Finally, the reward(s) consist of the outcome measures of the experiments.
\end{itemize}
The above list illustrates the generality of our approach: \pkg{StreamingBandit} can be used to allocate actions in all of the above applications. 

To ensure the computational scalability of \pkg{StreamingBandit} we assume that, at the latest interaction $t=t'$, all the information necessary to choose an action can be summarized using a limited set of parameters denoted $\theta_{t'}$, the dimensionality of $\theta_t$ often being (much) smaller than that of $\mathcal{D}_{t-1}$. Given this assumption, we identify the following two steps of a policy:
\begin{enumerate}
\item The decision step: In the decision step, using $x_{t'}$ and $\theta_{t'}$, and often using some (statistical) model relating the actions, the context, and the reward, which is parametrized by $\theta_{t'}$, the next action $a_{t'}$ is selected. Making a request to \pkg{StreamingBandit}'s \code{getaction} REST endpoint returns a JSON object containing the selected action. Optionally, the probability $p_{t'}$ of selecting this action (the \code{propensity}) and/or an identifier for this specific request (the \code{advice_id}), both of which are explained in more detail below, is also returned.
\item The summary step: In each summary step $\theta_{t'}$ is updated using the new information $\{x_{t'},a_{t'},r_{t'}, p_{t'}\}$. Thus, $\theta_{t'+1} = g(\theta_{t'}, x_{t'},a_{t'},r_{t'}, p_{t'})$ where $g()$ is some update function. Effectively, all the prior data, $\mathcal{D}_{t-1}$ are summarized in $\theta_{t'}$. This choice means that the computations are bounded by the dimension of $\theta$ and the time required to update $\theta$ instead of growing as a function of $t$. Note that this effectively forces users to implement an online policy \citep{Michalak2012} as the complete dataset $\mathcal{D}_{t-1}$ is not revisited at subsequent interactions. Making a request to \pkg{StreamingBandit}'s \code{setreward} endpoint containing a JSON object including either the \code{advice_id} or a complete description of $\{x_{t'},a_{t'}, p_{t'}\}$, and the reward $r_{t'}$, allows one to update $\theta_{t'+1}$ and subsequently to influence the actions selected at $t'+1$.
\end{enumerate} 

For the basic usage of \pkg{StreamingBandit} the experimenter---or rather an external server or mobile application---sequentially executes requests to the \code{getaction} and \code{setreward} endpoints, and allocates actions accordingly. Using this setup, \pkg{StreamingBandit} can be used to sequentially select advertisements on webpages, for example, allocate research subjects to different experimental conditions in an online experiment, or sequentially optimize the feedback provided to users off a mobile eHealth application. We provide a number of practical examples in section \ref{sec:practice}.

\subsection{Related approaches}

Theoretically, contextual decision-making relates to a broad literature ranging from active learning  \citep[e.g.,][]{beygelzimer2010agnostic, hanneke2014theory} to the general setting of reinforcement learning \citep{sutton1998introduction, szepesvari2010algorithms}. The contextual MAB problem \citep{Dudik2011a, Li2010b,agarwal2014taming} we consider here is a specific instance of reinforcement learning: it is a problem that is well-studied both without contextual information \citep{berry1985bandit} and in numerous generalizations, such as the continuous bandit \citep{mandelbaum1987continuous} and bandits with dependencies \citep{pandey2007multi}. The current work also relates to recent discussions on offline policy evaluation \citep{dudik2012sample, langford2011doubly}, although it is distinct from the multi-world testing service presented by \citet{agarwal2016making} in its focus on running (adaptive) policies online versus the online collection of data combined with the offline evaluation of policies. The field is too large to be properly reviewed in this paper, and we refer the reader to \citet{schwartz2017customer} and the references therein for an accessible introduction and contemporary applications.

Here we narrow our discussion of related approaches to related software projects, which we split into the following four categories: i) software for A/B testing, ii) software for general (supervised) learning, iii) software for offline policy evaluation, and iv) software for (sequential) optimization. The first category relates to our current project in that A/B tests---or randomized experiments---are used in many fields to address (C)MAB problems: one devotes a (pre-set) number of interactions to random exploration, after which the best performing action is selected and further exploited. This approach has become standard in many web companies \citep{jiang2016framework}. A more advanced version, often referred to as ``multi-variate testing" runs many A/B tests in parallel, possibly exploiting a factorial structure between the actions. Several commercial systems, such as Google Analytics, provide A/B testing abilities \citep{googleanalytics},  Optimizely \citep{optimizely}, and Mixpanel \citep{mixpanel}.

An effective policy depends heavily on the ability to predict the next reward given a context. Once available, a (large) dataset of contexts, actions, and rewards constitutes a supervised learning problem. Many general supervised learning solutions have been developed recently, such as \pkg{CNTK} \citep{seide2016cntk}, \pkg{GraphLab} \citep{collet2016leveraging}, \pkg{GeePS} \citep{cui2016geeps}, \pkg{MLlib} \citep{meng2016mllib}, \pkg{TensorFlow} \citep{abadi2016tensorflow}, and \pkg{Minerva} \citep{reagen2016minerva}. Some of these, such as \pkg{Vowpal Rabbit} \citep{langford2011vowpal} and \pkg{Jubatus} \citep{hido2013jubatus}, explicitly include libraries implementing specific bandit policies, or evaluation methods for bandit policies on existing, offline, data sets. Specific software projects for offline policy evaluation, and hence the ability to evaluate policies on existing datasets, are also available \citep[see, e.g.,][]{komiyama2015optimal,timnugent,ntucllab}. Others have provided language-specific code libraries implementing different policies, although most of these efforts seem to be a) geared towards computer scientists and experienced developers and b) not focused on field deployment \citep[see][and the references therein]{CapGarKau12,bgalbraith,danisola}.

There are a number of platforms that allow for sequential optimization: Google Analytics \citep{googleanalytics}, for example, supports Thompson sampling \citep{Agrawal2011, Kaptein2014a, thompson1933likelihood}, which is a method for sequentially allocating visitors to different actions dynamically based on the observed outcomes. However, contextual knowledge is not included. Yelp \pkg{MOE} \citep{YelpMoe2014} is an open-source software package that implements optimization over a large parameter space via sequential A/B tests in which Bayesian optimization is used to compute parameters for the next best A/B test. Finally, the Decision Service \citep{agarwal2016making} implements a number of functionalities implemented by \pkg{StreamingBandit} using a similar formalization (the summary and decision steps). This software package focusses on continuously collecting data to update and deploy policies that are evaluated offline, whereas \pkg{StreamingBandit} focusses on evaluating (adaptive) policies online.

\subsection{An overview of streaming bandit API calls}

\begin{figure}[htp]
  \centering
    \includegraphics[width=\textwidth]{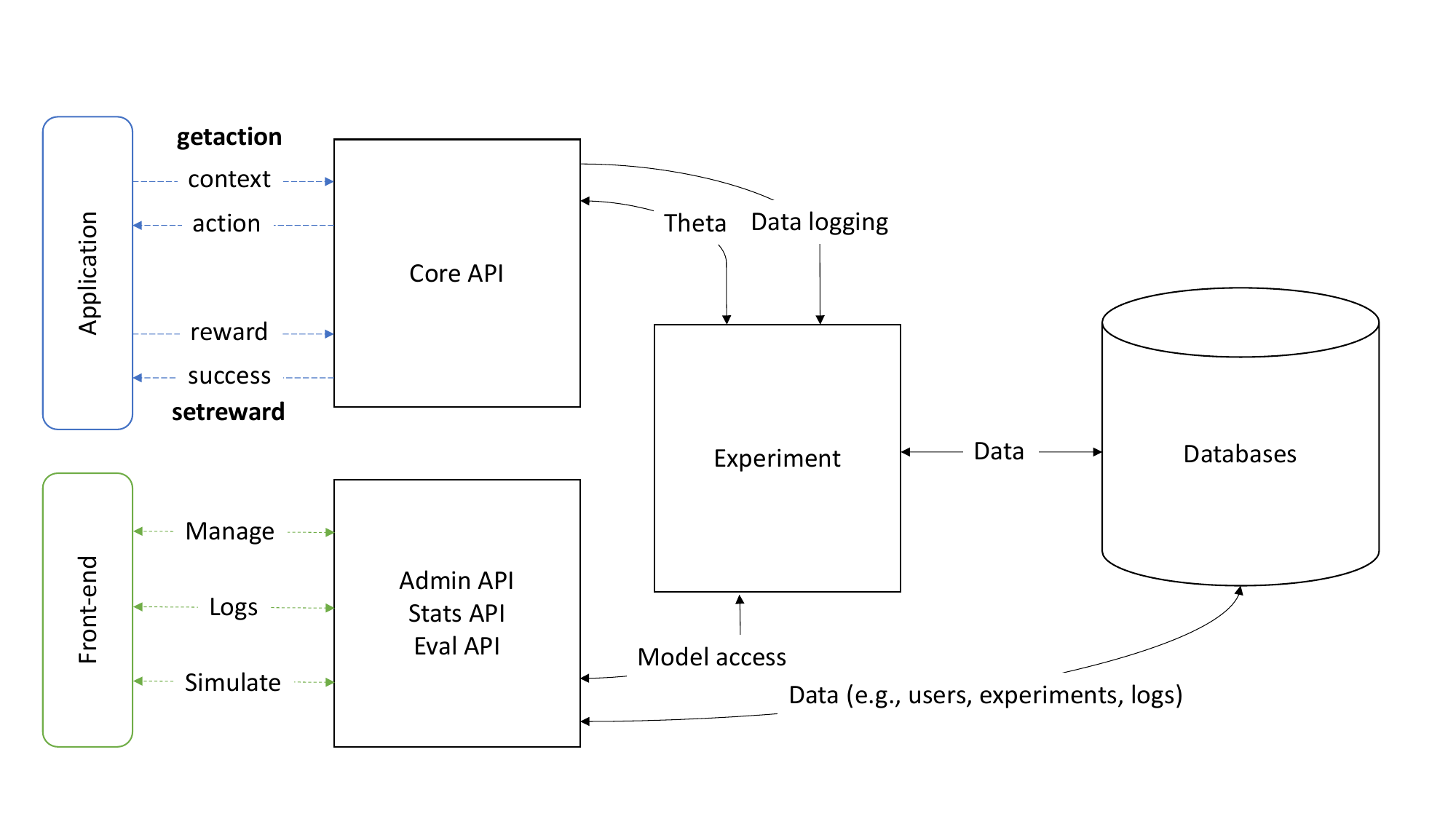}
    \label{fig:1}Python
      \caption{High-level architecture of StreamingBandit}
      \label{fig:architect}
\end{figure}

\pkg{StreamingBandit} is a \proglang{Python 3} application that runs a \pkg{Tornado} web server \citep{tornado} and discloses a REST API that facilitates the implementation of the summary and decision steps described above. A user of \pkg{StreamingBandit} first creates an experiment and subsequently implements---or adopts based on the library of available policies---a policy using \proglang{Python 3}. A policy specification consists of a) some code implementing the decision step given $\theta_{t'}$ and $x_{t'}$, and b) some code implementing the summary step given the observed outcomes to update $\theta_{t'}$. Figure \ref{fig:architect} presents an overview of the architecture of StreamingBandit. 

The application discloses a number of REST endpoints to facilitate the creation and editing of experiments and the extraction of data from running experiments. All endpoints apart from the \code{getaction} and \code{setreward} require the user to authenticate using a secure cookie. Logging in can be done by passing a JSON object to the \code{login} endpoint containing the parameters \code{username} and \code{password}; if the username and password are valid a secure- cookie is returned. New users can be created using the \code{user} call and posting the relevant information. For convenience, we provide a separate UI (a separate software project that can be found at \url{https://github.com/Nth-iteration-labs/streamingbandit-ui}) that allows easy point-and-click administration and management of experiments. Here we detail the primary endpoints and describe their functionality. We have already introduced the \code{getaction} and \code{setreward} calls, of which the full specification is:

\begin{itemize}
\item [GET] \code{getaction}: the query-string parameters consist of the experiment identification number, \code{exp_id} (string), a \code{key} (string), and the \code{context} (JSON). The call executes the decision step of a policy associated with the \code{exp_id} and returns an \code{action} (JSON), which optionally contains the elements \code{advice_id} (string), and \code{propensity} (float). The \code{key} is used to authenticate the request.
\item [GET] \code{setreward}: the query-string parameters consist of the \code{exp_id}, the \code{key}, the \code{reward} (JSON) and either the \code{advice_id}, in which case the \code{context} and \code{action} are retrieved from the associated \code{getaction} call, or the the \code{context} and \code{action} themselves. Subsequently, the summary step of the policy associated with the associated \code{exp_id} is executed and a JSON object containing the status is returned.
\end{itemize}

The primary endpoints at which to manage the experiments are: 

\begin{itemize}
\item [GET] \code{exp}: Returns a JSON object listing the \code{exp_id} and \code{name} of each experiment.
\item [POST] \code{exp}: Posting a JSON object containing the parameters \code{name}, \code{getcontext}, \code{getaction}, \code{getreward} and \code{setreward} creates a new experiment. The last four fields should contain executable restricted \proglang{Python 3} code. To ensure some safety in the executed code we limit the functionality of these customer scripts to a subset of \proglang{Python 3} code, using self-defined built-ins. This will disallow, for instance, the import of any other packages apart from the one we already make available. It also means that the user does not need to import any packages into the code because they are made available in the built-ins before any code is executed. The code in the \code{getaction} and \code{setreward} fields implements the decision and summary steps, respectively. The \code{exp} endpoint accepts a number of optional parameters, which we detail in section \ref{sec:advice-id}. A valid POST request to the \code{exp} endpoint returns a JSON object containing the \code{exp_id} and the \code{key} of the newly created experiment.

The code in the \code{getcontext} and \code{getreward} fields is not strictly necessary; these two snippets of code provide for the opportunity to simulate sequential decisions. This is extremely useful for debugging and can be used in simulation studies of a policy. Passing the query-string parameter \code{n} (int, default=1) to rest endpoint \code{eval/<exp-id>/simulate} sequentially executes the \code{getcontext}, \code{getaction}, \code{getreward} and \code{setreward} code of the associated experiment $n$ times.
\item [PUT] \code{exp/<exp_id>}: If the \code{exp_id} string in the url is a valid experiment for the current user, this call edits the existing experiment. The parameters are the same as those used for creating experiments.
\item [GET] \code{exp/<exp_id>}: Returns the name and \code{getaction} and \code{setreward} code for a specific experiment.
\item [DELETE] \code{exp/<exp_id>}: Deletes an experiment. When an experiment is deleted all the user-generated settings are removed, as well as the current $\theta$. However, logged data associated with the experiment is maintained.
\item [GET] \code{exp/<exp_id>/resetexperiment}: Resets the experiment: the current state of $\theta$ is deleted, but all the other information is retained and the policy can still be executed.
\end{itemize}

Next to these administrative calls, the application provides a number of calls to monitor running experiments and retrieve logged data. 

\begin{itemize}
\item [GET] \code{stats/<exp_id>/currenttheta}: Returns the current $\theta$ for the experiment as a JSON object.
\item [GET] \code{stats/<exp_id>/summary}: Returns an overview of the number of requests to the \code{getaction} and \code{setreward} endpoints.
\item [GET] \code{stats/<exp_id>/rewardlog}: Returns the logged \code{setreward} events (including the \code{context}, \code{action}, and \code{reward} objects) for the current experiment. It can be used for offline policy evaluation \citep[see, e.g., ][]{Li2010a, agarwal2016making}. The \code{limit} (int) query- string parameter limits the dump to the last $k$ events.
\item [GET] \code{stats/<exp_id>/actionlog}: Returns all the \code{getaction} events for the current experiment. Again, the \code{limit} parameters limit the dump to the last $k$ events.
\item [GET] \code{stats/<exp_id>/log}: Returns a JSON file of all data that was explicitly logged by the user using \code{self.log()} in the policy specification of an  experiment.
\end{itemize}

Requests made to non-existing REST endpoints result in a 404 status error, whereas erroneous calls to existing end-points return a JSON object containing a key \code{error} with an informative error message.

\subsubsection{Implemented policies: ``defaults''}
\label{sec:defaults}

\pkg{StreamingBandit} comes with a number of implemented policies to tackle standard (contextual) decision problems. A JSON object containing a list of defaults can be retrieved using the endpoint \code{default}, and the code for a specific default is obtained by calling \code{default/<default_id>}. We have implemented the following policies, amongst others:

\begin{itemize}
\item $\epsilon$-first: Implements the standard randomized clinical trial approach to the (C)MAB problem: the first $t < n$ interactions, where $n$ is set by the user, are allocated to actions randomly, after which the action with the highest average reward is selected for the remaining interactions.
\item $\epsilon$-greedy: Implements a greedy policy in which a proportion $p$ of interactions is randomly allocated to the available actions, whereas a proportion of $(1-p)$ interactions is allocated to the action with the highest average reward at that point in time.
\item Thompson sampling for the $k$-armed Bernoulli bandit: Thompson sampling provides a Bayesian solution to the MAB problem \citep{thompson1933likelihood, Agrawal2011}. We implement Thompson sampling for the Bernoulli bandit (e.g., $r \in \{0, 1\}$). Thompson sampling allocates actions proportional to one's current belief---as quantified using a posterior distribution---that an arm is optimal \citep{Kaptein2014a}. 
\item Lock-in Feedback: Lock-in Feedback is an allocation scheme for dealing with continuous actions ($a \in \mathbb{R}$) in which small systematic oscillations in the action choice over time are used to derive the gradient of the reward function and take a step toward the (local) maximum of that function \citep[see][for details]{kaptein2016tracking, kaptein2016investigation}.
\item Bootstrap Thompson sampling: Bootstrap Thompson sampling provides a computationally appealing alternative to Thompson sampling in cases in which it is hard to sample directly from the posterior distribution of a model online \citep[see][]{Kaptein}. In essence, the posterior distribution is approximated using an online bootstrap distribution \citep{Owen2012}.
\end{itemize}
We provide examples of the use of these policies in Section \ref{sec:examples}. \pkg{StreamingBandit} is easily extended and new defaults can be added by adding to the \code{/resources/defaults} folder of the application a folder with an informative name that contains the following four files:
\begin{enumerate}
\item \code{get_context.py}: A \proglang{Python} script that generates a JSON object encoding a context. 
\item \code{get_action.py}: A script that takes a JSON object encoding the context, and returns a JSON object containing the action.
\item \code{get_reward.py}: A script that generates a reward using a context and action JSON.
\item \code{set_reward.py}: A script that takes a context, action, and reward JSON and handles the logic of updating $\theta$.
\end{enumerate}
Restarting the web application after adding these files will automatically include the novel policy in the list of defaults. We welcome submissions of new default policies and other implementations. See section \ref{subsec:dev} (\textit{Further development}) for more details.

\subsubsection{StreamingBandit Libraries}
\pkg{StreamingBandit} was created to quickly create and test alternative policies in the field. This can be done by altering the \code{getaction} and \code{setreward} codes associated with an experiment. However, given that a number of operations are often encountered in the online processing of incoming data, \pkg{StreamingBandit} also provides a number of Python modules:
\begin{itemize}
\item \code{base}: This module provides functionalities for online (row-by-row) updates of (e.g.,) counts, means, variances, proportions, and covariances.
\item \code{lm}: Implements an online version of a linear regression model.
\item \code{bts}: Takes a model (e.g., \code{lm}) and a row of data and produces (or updates) an online bootstrap distribution of the parameters.
\item \code{lif}: Implements the Lock-in Feedback policy, as described in \citep{Kaptein2014d}.
\item \code{thompson}: Implements Thompson sampling for the $k$- armed Bernoulli bandit, amongst others.
\item \code{thompson_bayes_linear}: Implements model-based Thompson sampling using a Bayesian linear regression model.
\end{itemize}
New modules can be added to the application by adding a script to \code{/libs}. For detailed documentation of the individual modules we refer the reader to \url{http://nth-iteration-labs.github.io/streamingbandit/libs.html}.

\subsection{Installation, deployment, and documentation}

The \pkg{StreamingBandit} source code is available  from
\url{https://github.com/Nth-iteration-labs/streamingbandit/} and the documentation can be accessed on \url{http://nth-iteration-labs.github.io/streamingbandit/}. There are several ways in which \pkg{StreamingBandit} can be used:
\begin{enumerate}
\item At \url{http://sb.nth-iteration.com} we provide a running instance of \pkg{StreamingBandit}. You apply for a user account by sending an email to the corresponding author of this paper, and use our hosted webserver for (small-to-medium-sized) projects.
\item The easiest way to get going independently is probably to use our \pkg{Docker} container \citep{docker}. The following commands assume that you have \pkg{docker} and \pkg{docker-compose} installed, and that you are inside a folder in which you wish to put the source-code of \pkg{StreamingBandit}\footnote{For more information on how to get started with \pkg{Docker}, see \url{https://docs.docker.com/get-started/}.}. If so, starting
\pkg{StreamingBandit} requires, first, pulling the repository to your local system and going inside the folder:
\begin{Code}
$ git clone http://github.com/Nth-iteration-labs/streamingbandit/
$ cd streamingbandit
\end{Code}
Next, once you are inside the folder with all the source code, we can launch \pkg{StreamingBandit} by running:
\begin{Code} 
$ docker-compose up -d
$ docker exec -t streamingbandit_web_1 python3 ../insert_admin.py -p 
> test
\end{Code}
The first command makes sure that all necessary containers, including the databases, are running. The second command creates a user account admin with the password "test". To gracefully stop and start the container after running the first command, run the following command:
\begin{Code}
$ docker-compose stop
$ docker-compose start
\end{Code}
Starting the service will make \pkg{StreamingBandit} available at \url{http://localhost:8080} or the \pkg{Docker}-set IP address.

Note that the above commands only start the back-end REST service. The following commands are also needed to launch our front-end:
\begin{Code}
$ docker-compose -f docker-compose.yml -f docker-compose.front-end.yml 
> up -d
$ docker exec -t streamingbandit_web_1 python3 ../insert_admin.py -p 
  test
\end{Code}
The start and stop commands now change slightly as well:
\begin{Code}
$ docker-compose -f docker-compose.yml -f docker-compose.front-end.yml 
> stop
$ docker-compose -f docker-compose.yml -f docker-compose.front-end.yml 
> start
\end{Code}
which starts and stops both the front-end and the back-end at the same time. The front-end can be reached at \url{http://localhost} or the \pkg{Docker}-set IP address.

The front-end source-code can be found in a separate repository at \url{https://github.com/Nth-iteration-labs/streamingbandit-ui}, but for this use-case it is not necessary to download the repository to your local system because we have uploaded a \pkg{Docker} image to the internet and \pkg{Docker} will download that image automatically via the \code{docker-compose} command.
\item For larger-scale projects we recommend installing from source and perhaps using a load-balancer. For details, please consult the documentation at \url{http://nth-iteration-labs.github.io/ streamingbandit/}.
\end{enumerate}

\subsubsection{Further development}
\label{subsec:dev}

The above sections give the essential details of \pkg{StreamingBandit}. We gladly accept any contributions towards making \pkg{StreamingBandit} better and more useful. The guidelines for contributing to the development of \pkg{StreamingBandit} can be found in the documentation. 



\section{Getting Started}
\label{sec:usage}

In the remainder of this article we assume that the reader is running the default \pkg{Docker} container installation of \pkg{StreamingBandit}, and is using the management front-end for the administration of experiments. In introducing the details of setting up a policy we describe the setup and usage of a simple---but very frequently used---policy: $\epsilon$-first. When this policy is executed a sample of size $n$ interactions is uniformly randomly allocated to a control ($a = \text{control}$) or treatment ($a = \text{treatment}$) action (or condition), after which the treatment is adopted if it is more effective than the control condition. With slight abuse of the notation this can be denoted:

\begin{eqnarray}
\label{eq:pol}
\pi_{\epsilon\texttt{-first}}(\mathcal{D}, n) = 
\begin{cases}
a_t \sim \texttt{random}(\text{control}, \text{treatment}) & \text{if } t \leq n \\
a_t = \texttt{max}( \bar{r}_{control}, \bar{r}_{treatment} ) &\text{otherwise}
\end{cases}
\end{eqnarray}
where $\bar{r}_{control}$ denotes the sample average of outcomes observed in the control condition when $t \leq n$, and the last line denotes selection of the action with the highest empirical average reward when $t>n$. 

\begin{figure}[htp!]
  \centering
    \includegraphics[width=.85\textwidth]{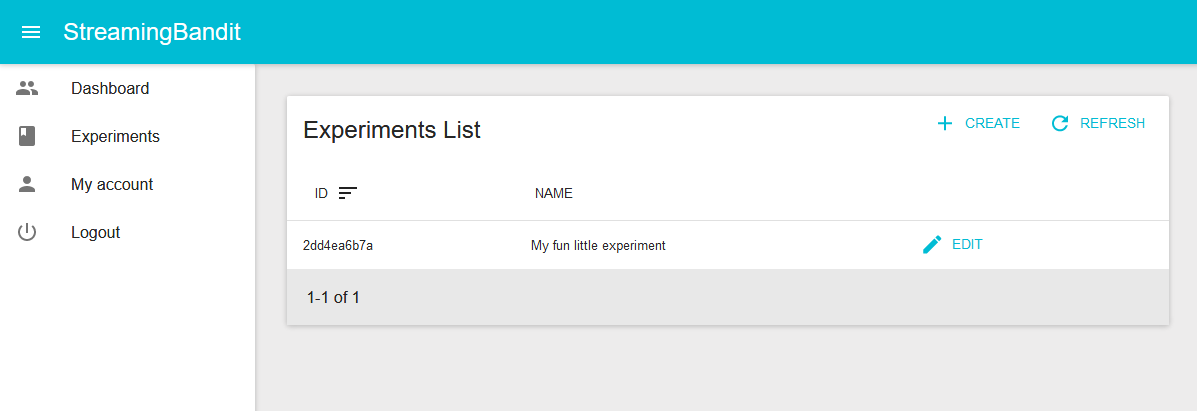}
    \label{fig:front-end}
      \caption{Screen shot of the default front-end for \pkg{StreamingBandit}.}
\end{figure}

The management front-end---of which Figure \ref{fig:front-end} shows  a screenshot---makes it easy to create a new experiment or to use  one of the defaults as a starting point for creating one's own policies. We present the front-end in more detail in the Appendix. Once the experiment has been created it receives an \code{exp_id} and a key \code{key}. This enables the REST endpoints
\begin{Code} 
http://HOST/getaction/<exp_id>?key=<key>&context={}
\end{Code}  
and
\begin{Code} 
http://HOST/setreward/<exp_id>?key=<key>&context={}&reward={}&action={}.
\end{Code} 

The actual functionality is provided by the \code{getaction} and \code{setreward} code specified when the experiment is created, whereas the \code{getcontext} and \code{getreward} codes are useful for simulations and testing. Below we detail each of these in turn for the version of $\epsilon$-first implemented in the defaults. Before that we should note that we will denote a few variables and functions using \code{self} inside the code. These variables and functions are denoted with \code{self} because they are part of the experiment class in which the custom code runs. For the most part, we will only use a reference to \code{self} with the following variables and functions:

\begin{itemize}

\item \code{self.context}
\item \code{self.action}
\item \code{self.reward}
\item \code{self.get_theta()}
\item \code{self.set_theta()}
\end{itemize}

The code for a simple $\epsilon$-first implementation is as follows:

\begin{itemize}

\item \code{getcontext}: The canonical $\epsilon$-first strategy does not consider a context. Hence, we leave this blank.
\item \code{getaction}: The implementation of the decision step of $\epsilon$-first is:
\begin{Code}
n = 100
mean_list = base.List(
             self.get_theta(key="treatment"), 
             base.Mean, ["control", "treatment"]
             )
if mean_list.count() >= n:
    self.action["treatment"] = mean_list.max()
    self.action["propensity"] = 1
else:
    self.action["treatment"] = mean_list.random()
    self.action["propensity"] = 0.5
\end{Code}
This code uses a number of libraries implemented in \pkg{StreamingBandit}: below we detail each line in turn. First, the sample size of the experiment, $n$ in Equation \ref{eq:pol}, is set. The next line of code generates a list of \code{base.Mean} objects. This object provides the functionality to compute streaming updates of sample averages, and the list contains one such average for each of the possible treatments specified by name, using \code{["control", "treatment"]}. The \code{self.get\_theta()} call is used to retrieve $\theta_{t'}$, which in this case thus contains two \code{base.Mean} objects named ``control" and ``treatment". A count, $n$, and mean reward, $\bar{r}$, are contained within each \code{base.Mean}  object.

The resulting \code{mean_list} object thus, in this case, contains two \code{base.Mean} objects, each of which contains a mean value and a count that can be updated and manipulated. In the next lines the total count of the number of observations over all mean elements in the list is retrieved. If this is larger than $n$, the treatment with the highest average value is returned, otherwise a random element of the list is returned. The returned JSON object when making a call to \url{http://HOST/<exp_id>/getaction?key=<key>} and filling in the correct \code{exp_id} and \code{key} appears as follows:
\begin{Code}
{"action": 
     {"treatment": "control", 
     "propensity": 0.5},
"context": {}}
\end{Code}
where the value of \code{treatment} changes randomly as long as $n \leq t$.

\item \code{getreward}: Rewards can be simulated by using a few lines of \proglang{Python 3} code
\begin{Code}
if self.action["treatment"] == "control":
    self.reward["value"] = np.random.normal(4, 1)
else:
    self.reward["value"] = np.random.normal(6, 2)
\end{Code}
in which the rewards for the selected action in the decision step are drawn from a normal distribution ($r_{control} \sim \mathcal{N}(4, 1)$, $r_{treatment} \sim \mathcal{N}(6, 2)$).

\item \code{setreward}: When a reward has been generated, the summary step for the $\epsilon$-first policy is implemented as:
\begin{Code}
n = 100
mean_list = base.List(
    self.get_theta(key="treatment"), 
    base.Mean, ["control", "treatment"]
    )

if mean_list.count() < n:
    mean = base.Mean(
         self.get_theta(
         key="treatment", value=self.action["treatment"])
         )
    mean.update(self.reward["value"])
    self.set_theta(
         mean, key="treatment", 
         value=self.action["treatment"]
         )
\end{Code}
which again uses the \code{libs.base} library. After this the action is retrieved and the associated mean object is updated using \code{mean.update} as long as the exploration phase is ongoing. The last line stores $\theta_{t'+1}$ such that it can be retrieved again for future decision-making. In this implementation, after the experiment when $n>t$, $\theta$ is no longer updated. Note that a slightly more elaborate version of this example that facilitates propensity scores (see \ref{sec:propensity}) can be found in the defaults (see
\ref{sec:defaults}).

\end{itemize} 

As stated above, the \code{getcontext} and \code{getreward} codes are not strictly necessary to use the implemented policy in field  studies; these two snippets of code merely provide the opportunity to simulate an experiment, a feature that is extremely useful for debugging. In actual evaluations of a policy the data resulting from these calls would be sent by the outside world (e.g., via a website or mobile application). However, to demonstrate the utility of the \code{getcontext} and \code{getreward} codes, note that a request to the endpoint \code{/eval/<exp_id>/simulate} with parameters \code{N=150}, \code{seed=1271246}, and \code{verbose=False} yields the following JSON response:
\begin{Code}
{
    "theta": {
        "treatment:control": {
            "n": "52",
            "m": "4.0259030511640885"
        },
        "treatment:treatment": {
            "n": "48",
            "m": "5.829777419810004"
        }
    },
    "simulate": "success",
    "experiment": "121e3e0aeb"
}
\end{Code}
which shows the number of times the \code{treatment} and \code{control} conditions were selected (\code{n}) and their respective mean reward (\code{m}). Although we simulated $150$ interactions, the total number of interactions stored in $\theta$ is $48 + 52 = 100$ because in the implementation above we stop updating $\theta$ when $t>n$.

\subsection{Additional features}

We described the setup and simulation of a simple bandit experiment in the previous section. The description skipped over a number of useful features of \pkg{StreamingBandit}, which we address below.

\subsubsection{Offline analysis of bandit policies}
\label{sec:propensity}

When we first introduced the \code{getaction} endpoint we mentioned the optional return field \code{propensity}. In a number of default policies, the return object contains this propensity $p_t$, which is the probability of selecting the action at interaction $t$. By way of an illustration, for $\epsilon$-first, as detailed above, the computation of $p_t$ is as follows:
\begin{eqnarray}
\label{eq:pol}
p_t = 
\begin{cases}
.5 & \text{if } t \leq n \\
1 & \text{otherwise}
\end{cases}
\label{simple:experiment}
\end{eqnarray}

Whenever it is possible to compute these propensities---which is sometimes difficult, such as when $a \in \mathbb{R}$---the default policies include $p_t$. This serves two purposes:
\begin{enumerate}
\item When addressing contextual sequential decision problems, and when the probability of selecting an action depends on the context, the propensity $p_t$ can be used for inverse propensity matching or weighting \citep{austin2011introduction} to improve the estimate of the causal effect of the action by accounting for the contextual covariates \citep[see, e.g.,][]{guido2015causal, pearl2009causality}.
\item When $p_t$ is included, the logged data of an experiment can be used for the offline evaluation of alternative decision policies. This can be attained by using inverse propensity scoring (ips). Suppose we are evaluating a policy $\pi$ using a logged dataset containing $N$ events. The ips estimate of average reward of the policy can be obtained by computing
\begin{eqnarray}
\texttt{ips}(\pi) & = & \frac{1}{N} \sum_{t=1}^{N} \mathbbm{1} \{ \pi(x_t) = a_t \} r_t / p_t
\end{eqnarray}
where the indicator is $1$ when the action of $\pi$ matches the action in the logs. \citet{agarwal2016making} provide a more extensive discussion of the benefits of using offline methods to evaluate alternative policies.
\end{enumerate}

\subsubsection{Advice ID, delayed rewards, and logging}
\label{sec:advice-id}

When we described the [POST] \code{exp} endpoint we omitted a number of optional parameters that can be supplied in the JSON object. First of all, we skipped discussion of the \code{advice_id} parameter. This Boolean indicates whether or not the \code{getaction} call should return an \code{advice_id}. When set to \code{True} the \code{advice_id} parameter enforces a direct link between the \code{getaction} and \code{setreward} endpoints. In the example discussed above we were implicitly assuming that the application consuming the REST API would handle the logic that ensures that by the time the \code{setreward} endpoint is called, the \code{context}, \code{action} (including the \code{propensity}), and \code{reward} are properly supplied. However, this could be challenging for some consuming applications. In such cases, setting \code{advice_id = True} would require the consuming application to merely specify the \code{advice_id} when making a request to the \code{setreward} endpoint; \pkg{StreamingBandit} will merge the actions and context that were provided earlier in the associated \code{getaction} call with the rewards supplied in the \code{setreward} call.

When setting \code{advice_id = True}, one can also specify a) how  long the \code{advice_id} will be retained (in hours). This is useful in some specific applications. In an online advertising experiment, for example, when a click on an advertisement is not registered within $12$ hours it is extremely unlikely that this will happen in the future; it is more likely that the appropriate call to the \code{setreward} with $r_t = 0$ failed to register. Setting \code{delta_hours=12} and \code{default_reward=\{"reward":"0"\}} ensures that after twelve hours the \code{setreward} call associated with the \code{advice_id} is automatically executed with a reward of zero. It should also be noted that although all the examples provided in this paper sequentially execute the \code{getaction} and \code{setreward} calls, this is not at all a necessity. However, any bias in a (learning) model that might originate from (e.g.,) a delay in the arriving data in the \code{setreward} calls should be explicitly handled by the user.

Finally, we have not yet discussed the \code{hourly_theta} Boolean: if this is set to \code{True} when creating the experiment, the state of $\theta$ will be logged every hour. Calling \code{stats/<exp_id>/ hourlytheta} with parameter \code{limit} returns the last $k$ of these snapshots of $\theta$, which could be useful for monitoring the progress of an experiment over time.

\subsubsection{The nesting of policies}

In addition to the libraries described earlier, and the \code{self.get_theta()} and \code{self.set_theta()} methods for storing and retrieving data, there are a number of methods available to the user from the code supplied in the \code{getaction}  and \code{setreward} fields. The most interesting of these is the ability to instantiate other experiments within a running experiment. By way of illustration, the code
\begin{Code}
experiment = Experiment(exp_id =  <exp_id>)
self.action = experiment.run_action_code(context =  self.context)
\end{Code}
can be used to run the \code{getaction} code of the experiment with \code{exp_id=<exp_id>} from another experiment. Similarly, \code{experiment.run_reward_code()} would execute the \code{setreward} code for another experiment. This allows the user to nest different experiments, and hence to essentially use a sequential decision policy $\pi^*$ to decide from among a range of policies that are being executed $\pi_{1, \dots, k}$. We provide a working example of this policy nesting in the section
\ref{sec:nesting}.

\subsection{Performance}

To examine the performance of our RESTful API we set up an Ubuntu 16.04 x64 quad-core virtual server with 16GB of RAM running the \pkg{StreamingBandit} server, and additionally installed the \pkg{wrk2} load generator on a smaller (single core, 1GB RAM) Ubuntu 16.04 x64 machine connected to the same subnet within the same datacenter. We chose Wrk2 \citep{tenetene} as our load generator, as it is a HTTP benchmarking tool that is capable of generating significant load when run on a single CPU, and can easily be extended to test different RESTful HTTP methods through the use of \proglang{Lua} scripts.

To ensure that our load tests would not be hampered by OS related limitations we optimized sysctl.conf on both machines, turning off disk swapping, upping the number of connections per port, and optimizing port reuse. We also tested our client-server throughput with \pkg{iPerf3} \citep{Iperf.fr2016}. These tests indicated a throughput of 736 Mbits/s - more than enough bandwidth to safeguard against system-level I/O bottlenecks interfering with our API-level tests.

On completion of our test-bed we proceeded to run several \pkg{wrk2} load tests, focusing on industry-standard API performance measures \citep{De2017}. The results for a single \pkg{wrk2} thread running 100 concurrent AB test getaction calls at a time with a throughput limit of 1000 requests per second were the following:

\begin{itemize}
  \item Average, max and standard deviation of latency: 21.09ms, 90.56ms, 13.22ms
  \item Throughput, in requests per second: 100 (equal to max set wrk2 throughput)
  \item Top total CPU utilization: 69\% (Of which: Python3 65\% of one of four available CPU's)
  \item Top Heap memory utilization: 3\%
\end{itemize}

\begin{figure}[htp!]
  \centering
    \includegraphics[width=.85\textwidth]{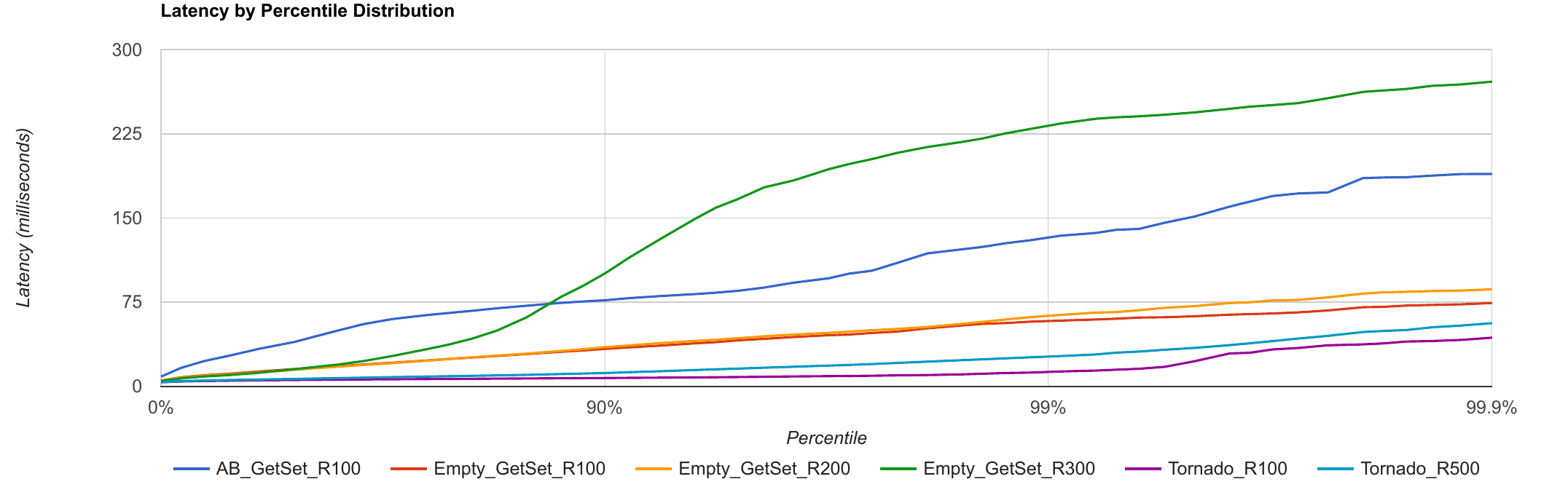}

      \caption{Latencies of basic \pkg{Tornado} calls when taxed by \pkg{wrk2} at a maximum throughput of 100 (Tornado\_R100) versus 500 (Tornado\_R500) calls per second (cps), as compared to \pkg{StreamingBandit} "AB test" (AB\_GetSet\_R100, throughput limited at 100 cps) and empty \code{getaction}/\code{setreward} calls (with Empty\_GetSet\_100, Empty\_GetSet\_200, Empty\_GetSet\_300 at respectively 100, 200 and 300 cps).}
    \label{fig:benchmarks}
\end{figure}

When we compared these numbers against some representative \proglang{Python} web framework benchmarks \citep{Klenov2015} we found that \pkg{StreamingBandit} could hold its own. Still, to obtain a more objective measure of how "empty" versus "AB test" StreamingBandit \code{getaction} calls measure up to basic, vanilla \pkg{Tornado} requests, we compared these as well. The results, as illustrated in Figure \ref{fig:benchmarks}, demonstrate that \pkg{StreamingBandit} adds little overhead to basic \pkg{Tornado} processing, and scales well up to 250 to 300 requests per second when running on a single virtual CPU core. The relatively minor increment in throughput and latency between the "empty" and the "AB test" experiments further indicates that \pkg{StreamingBandit} offers sufficient capacity to implement more complex experiments.

\section{Examples of the implemented policies}
\label{sec:examples}

In the following we work out a number of different (C)MAB policies. First, we present a simple implementation of $\epsilon$-greedy \citep{sutton1998introduction}, then we introduce Thompson sampling for the canonical $k$ armed Bernoulli bandit \citep{thompson1933likelihood}, and for optimal design in between-subject experiments \citep{Kaptein2014a}. We proceed by demonstrating two possible policies to deal with the continuum-bandit problem (problems in which $a \in \mathbb{R}$): Bootstrap Thompson sampling for a CMAB problem using a simple linear model \citep{Kaptein} and Lock-in Feedback (LIF) \citep{Kaptein2014d}. We further demonstrate how \pkg{StreamingBandit} can be used to nest multiple policies, and show how \pkg{StreamingBandit} can be used to evaluate multiple policies in parallel using the offline evaluation method proposed by \citet{Li2010a}. This latter approach is, to the best of our knowledge, novel. All of the implementations discussed in this section can be found in the defaults (see \ref{sec:defaults}).

\subsection{E-greedy}
\label{subsec:egreedy}

One frequently used policy is called $\epsilon$-greedy \citep{sutton1998introduction}. It is implemented in a simple problem consisting of a control and a treatment arm, as we considered when we introduced $ \epsilon$-first, by playing the arms uniformly randomly with some probability $\epsilon$, and selecting the hitherto best-performing arm with probability $1-\epsilon$. The same \code{getcontext} and \code{setreward} codes as in our $\epsilon$-first example above are used to implement $\epsilon$-greedy, as follows:

\begin{itemize}
\item \code{getaction}: 
\begin{Code}
e = .1
mean_list = base.List(self.get_theta(key="treatment"), 
    base.Mean, ["control", "treatment"]
    )
if np.random.binomial(1,e) == 1:
    self.action["treatment"] = mean_list.random()
    self.action["propensity"] = 0.1*0.5
else:
    self.action["treatment"] = mean_list.max()
    self.action["propensity"] = (1-e)
\end{Code}
Where, contrary to our $\epsilon$-first example, we explicitly include the computation of the propensity $p_t$.
\item \code{setreward}: The summary step for the $\epsilon$-first can be implemented as:
\begin{Code}
mean = base.Mean(self.get_theta(
         key="treatment", value=self.action["treatment"]
         ))
mean.update(self.reward["value"])
self.set_theta(mean, key="treatment", value=self.action["treatment"])
\end{Code}
which is the same as for $\epsilon$-greedy except for the fact that the respective means are updated at each interaction $t$ instead of $n<t$.
\end{itemize}

Running a simulation with $n=1000$ and $\texttt{seed}=1271246$ gives:
\begin{Code}
{
    "theta": {
        "treatment:control": {
            "n": "70",
            "m": "4.011771491758239"
        },
        "treatment:treatment": {
            "n": "930",
            "m": "5.9609857188253015"
        }
    },
    "simulate": "success",
    "experiment": "3ea45886b5"
}
\end{Code}
in which it is clear that the \code{treatment} arm is preferred.

\subsection{Thompson sampling for the $K$-armed Bernoulli bandit}
\label{sec:thompson}

As our second example we provide the code to implement Thompson sampling for the classical Bernoulli bandit problem where the rewards are either $0$ or $1$, and for each arm $k=1, \dots, K$ the probability of success (reward = 1) is $\mu_k$ \citep{Kaufmann2012}. Thompson sampling is a Bayesian policy in which one selects an action with a probability that is  proportional to one's posterior belief that the action is optimal \citep[see][for details]{Kaufmann2012}. In the Bernoulli reward case the $\text{Beta}(\alpha,\beta)$ distribution provides a convenient a priori choice in that after observing a Bernoulli trial the posterior distribution is simply $\text{Beta}(\alpha+1,\beta)$ in the case of success, and $\text{Beta}(\alpha,\beta+1)$ in the case of failure. Using $S_k$ and $F_k$ to denote the number of failures and successes for arm $k$, both of which are $0$ at the start, Thompson sampling proceeds as follows; at each interaction $t$,
\begin{enumerate}
\item for each arm $k=1, \dots, K$, sample $d_k(t)$ from $\text{Beta}(S_k+1,F_k+1)$,
\item select arm $k(t) = \argmax_{k} d_k(t)$,
\item and if $r_t = 1$ then $S_k = S_k+1$ or when $r_t = 0$ then $S_k = S_k+1$.
\end{enumerate}

Thompson sampling for the $4$-arm Bernoulli bandit problem can be implemented as follows:
\begin{itemize}
\item \code{getcontext}: The Bernoulli bandit does not consider a context; we leave this field blank.
\item \code{getaction}: The decision step, using the \code{libs.thompson} library, can be implemented using:
\begin{Code}
propl = thmp.BBThompsonList(
     self.get_theta(key="treatment"), ["1","2","3","4"]
     )
self.action["treatment"] = propl.thompson()
self.action["propensity"] = propl.propensity(self.action["treatment"])
\end{Code}
where the four arms are indexed using the numbers $1-4$.

\item \code{getreward}: Bernoulli rewards can be simulated using:
\begin{Code}
self.reward["value"] = np.random.binomial(
     1,(0.2*int(self.action["treatment"]))
     )
\end{Code}
which produces Bernoulli rewards with a probability of $.2,.4,.6,.8$ for the four arms respectively.

\item \code{setreward}: Finally, the updates of the posterior distributions are implemented using
\begin{Code}
prop = base.Proportion(self.get_theta(
     key="treatment", value=self.action["treatment"])
     )
prop.update(self.reward["value"])
self.set_theta(prop, key="treatment", value=self.action["treatment"])
\end{Code}

\end{itemize}

Running a simulation with $n=1000$ and $\texttt{seed}=1271246$ gives:
\begin{Code}
{
    "theta": {
        "treatment:1": {
            "p": "0.14285714285714288",
            "n": "7"
        },
        "treatment:4": {
            "p": "0.8025404157043878",
            "n": "866"
        },
        "treatment:2": {
            "p": "0.20000000000000004",
            "n": "10"
        },
        "treatment:3": {
            "p": "0.5982905982905986",
            "n": "117"
        }
    },
    "experiment": "1de9753f51",
    "simulate": "success"
}
\end{Code}
Which demonstrates that arm $4$ is clearly, and correctly, preferred.

\subsection{Thompson sampling for optimal design}

Another example that could have practical relevance in social-science experiments is presented in \citet{Kaptein2014a}: when running an experiment comparing two groups that receive different treatments, assuming unequal variances in the observed continuous outcomes, it is beneficial to allocate a larger number of subjects to the treatment with the highest variance to increase the precision in the obtained effect-size estimate. The Thompson sampling policy to implement this sequential allocation is to compute---using a normal-inverse $\chi^2$ model---the posterior variances $\sigma^2_1$ and $\sigma^2_2$ of the two treatments in the summary step. Next, in the decision step, a draw $d$ from each of the two posterior distributions $\sigma^2_1$ and $\sigma^2_2$ is obtained and the treatment is selected for which $\frac{d}{n}$, where $n$ denotes the number of subjects allocated to the respective treatment, is highest. This choice leads to the largest reduction in the estimated standard error of the mean difference between the two groups. We refer the interested reader to \citet{Kaptein2014a} for details. This sequential allocation scheme can be implemented In \pkg{StreamingBandit} using:

\begin{itemize}
\item \code{getcontext}: Left blank as no context is considered
\item \code{getaction}: In the summary step, we retrieve a list of two variance objects one for each treatment. Variance objects, and the ability to update these online, are included in \code{base} library. Next, we implement Thompson sampling on the level of the posterior variances of the outcomes; this is included in the \code{libs.thompson} library:
\begin{Code}
varList = thmp.ThompsonVarList(
    self.get_theta(key="treatment"), ["control","treatment"]
    )
self.action["treatment"] = varList.experimentThompson()
\end{Code}
\item \code{getreward}: To simulate outcomes with unequal variances we can use:
\begin{Code}
if self.action["treatment"] == "control":
    self.reward["value"] = np.random.normal(0, 1)
else:
    self.reward["value"] = np.random.normal(1, 5)
\end{Code}
\item \code{setreward}: And finally, we update the respective posterior variance when new observations arrive:
\begin{Code}
var = base.Variance(self.get_theta(key=self.action["treatment"]))
var.update(self.reward["value"])
self.set_theta(var, key="treatment", value=self.action["treatment"])
\end{Code}
\end{itemize}

Running a simulation with $n=100$ and $\texttt{seed}=43123$ gives:

\begin{Code}
{
    "theta": {
        "treatment:treatment": {
            "s": "1453.3754330265062",
            "n": "77",
            "x_bar": "0.777831868342291",
            "v": "19.123360960875083"
        },
        "treatment:control": {
            "s": "32.31094303640007",
            "n": "23",
            "x_bar": "0.032257238191552844",
            "v": "1.4686792289272759"
        }
    },
    "experiment": "84b4d7eda",
    "simulate": "success"
}
\end{Code}
This result highlights two things: First, it is clear that the treatment condition with the highest variance is indeed selected more often. This is the expected behavior to ensure that the precision of the estimate is increased. Second, the result demonstrates the internals of the \code{base.Variance} object: to compute a variance in a data stream we maintain a count (\code{n}), a mean (\code{m}), and the numbers \code{s} and \code{v}; of these \code{v} is the current sample variance, whereas \code{s} in an auxiliary variable used to implement Welford's method for computing a variance online \citep{welford1962note}.

\subsection{Bootstrap Thompson Sampling}

Bootstrapped Thompson sampling (BTS) is a recent approach devised to address CMAB problems \citep[see,  e.g.,][]{Kaptein,Osband2015}. The basic idea behind BTS is that instead of using a draw from the posterior distribution of the parameters of interest to decide on the next allocation, as is the case in previous Thompson sampling examples, one can maintain, online, a number of bootstrapped estimates of the parameters. These bootstrapped estimates can then be used to balance exploration and exploitation by randomly selecting one of the bootstrap replicates \citep[see][for details]{Kaptein}.

\pkg{StreamingBandit} implements this sequential allocation scheme quite generally using the double-or-nothing bootstrap \citep{Owen2012}. The appeal of BTS compared to traditional Thompson sampling is that a) it can be fully carried out online as long as the point estimates of interest can be obtained online, and b) it can be used in many situations in which obtaining draws from the true posterior density of interest is computationally difficult. Here we provide a simple example of the implementation of BTS using a linear model to relate the actions, the contexts, and the rewards.

For ease of exposition, let us consider a practical example. Suppose we are concerned with choosing a \code{price} (the action) of a product sold online such that the \code{revenue} is maximized (the reward). Let us further assume that we believe the relation between these two quantities is quadratic, and that we think the optimal sales price differs between \code{new} customers and \code{returning} customers. The following code implements this scenario such that it can be simulated:

\begin{itemize}
\item \code{getcontext}: The get context code simulates the visit of either a new or a returning visitor.
\begin{Code}
self.context["customer"] = random.choice(["new", "returning"])
\end{Code}
\item \code{getaction}: Next, the get action code, which is slightly more involved, uses the \code{lm} library to instantiate $m=100$ linear models of the form
\begin{eqnarray}
\texttt{revenue}  \sim & \beta_0 + \beta_1 \texttt{price} + \beta_2 \texttt{price}^2 + \beta_3 \texttt{new} +  \\
				& \beta_4 \texttt{price} * \texttt{new} + \beta_5 \texttt{price}^2 * \texttt{new}.
\end{eqnarray}
Here, the starting values of the model $\beta$'s are initially set to zero. The \code{BTS} object maintains $m=100$ of these models, whereas the remaining code samples one of these $m=100$ models and computes the \code{price} that maximizes the expected revenue given the current customer and the current state of the parameters. We add comments to the code to improve readability:

\begin{Code}
# Instantiate BTS with m=100 samples:
BTS = bts.BTS(self.get_theta(), lm.LM, m = 100, default_params = \
       {'b': np.zeros(6).tolist(), 'A' : np.identity(6).tolist(), 'n' : 0})

# Return one of the m samples:
model = lm.LM(default = BTS.sample())

# Retrieve its coefficients:
betas = model.get_coefs()

# Create dummy for customer
if(self.context["customer"] == "returning"):
    customer = 1
else:
    customer = 0

# Maximize the function
if betas[2] != 0 or betas[5] != 0:
    x = ( (-(betas[1] + betas[4] * customer)) / 
        (2*(betas[2] + betas[5] * customer)) )
    x = np.asscalar(x)
    if x < 5 or x > 20:
        x = np.random.uniform(5,20)
else:
    x = np.random.uniform(5,20)

# Return the price
self.action["price"] = x
\end{Code}
Note that we restrict the prices to be between $5$ and $20$, such that if \code{BTS} needs some more exploration, it will not go towards extreme values, which may happen if a linear model is selected that has no parabola - in a field experiment you might want not have your prices restricted to certain ranges as well.

\item \code{getreward}: In the get reward code we use a logistic function to simulate the probabilities of accepting or rejecting the product at the offered price for different customer types.
\begin{Code}
# Get parameters
# Create dummy for customer
if(self.context["customer"] == "returning"):
    customer = 1
else:
    customer = 0
price = self.action["price"]

# Create logistic function
logistic = lambda x: 1 / (1 + numpy.exp(-x))

# Compute purchase yes / no
buy = numpy.random.binomial(1, logistic(-0.1 * (price - (10+4*customer))**2))

# Compute the reward
self.reward["revenue"] = buy * price
\end{Code}
Here it is clear that new customers are more inclined than returning customers to buy for higher prices, the revenue-maximizing price being $\approx 10.9$ for new customers, and $\approx 14.7$ for returning customers.

\item \code{setreward}: Finally, after generating the reward, the summary step for this policy can be implemented as follows:
\begin{Code}
# Extract values:
# Create dummy for customer
if(self.context["customer"] == "returning"):
    customer = 1
else:
    customer = 0
price = self.action["price"]

# Create feature vector and response:
X = [1, price, price**2, customer, customer*price, customer*price**2]
y = self.reward["revenue"]

# Instantiate the m=100 lm models
BTS = bts.BTS(self.get_theta(), lm.LM, m = 100, default_params = \
       {'b': np.zeros(6).tolist(), 'A' : np.identity(6).tolist(), 'n' : 0})

# Update the model parameters using the new observation
BTS.update(y, X)

# Store the updated values
self.set_theta(BTS)
\end{Code}
\end{itemize}

\begin{figure}[h!]
  \centering
    \includegraphics[width=.7\textwidth]{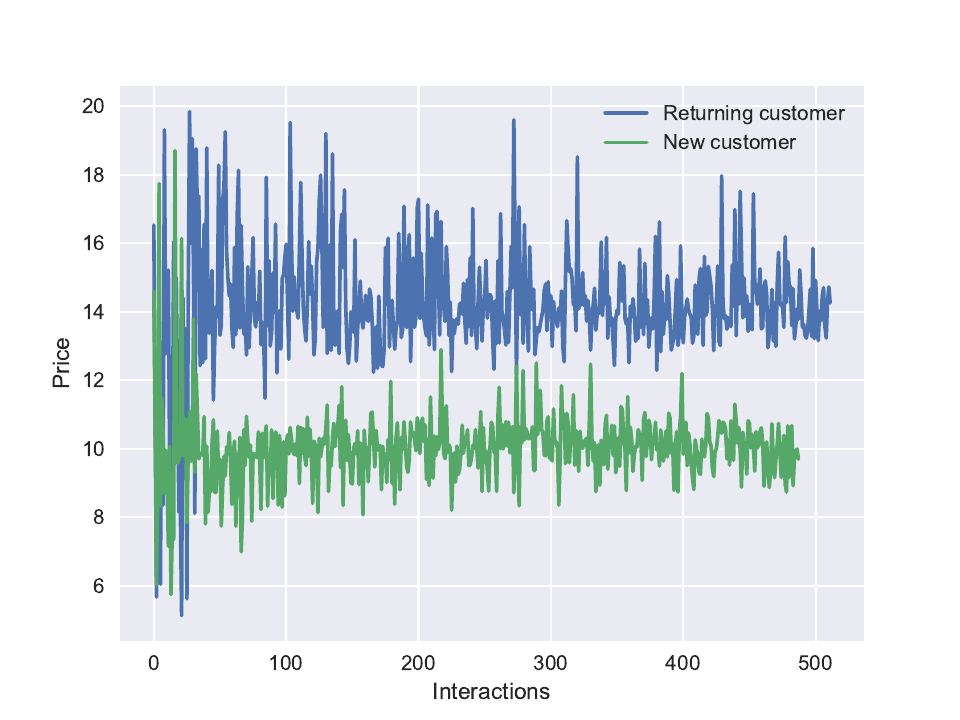}
      \caption{Overview of the selected prices of BTS with $m = 10$ and $N = 1000$ for both returning and new customers (separate lines).}
      \label{fig:bts}
\end{figure}

To illustrate the outcomes of this sequential allocation scheme we run a simulation with $N=1000$ and $\texttt{seed}=43123$ setting the ``log results'' to True. Next, using the logged data, we plot the selected prices for each of the customer types separately. Figure \ref{fig:bts} shows the progression of the recommended prices for each customer type; it is clear that these display a lot of exploration behavior early in the data stream, but after about $100$ observations the BTS policy seems to exploit more and settles on a price that is close to the maximum in a large number of the interactions.

\subsection{Lock-in Feedback}
\label{sec:lif}

Picking a price was considered the intended action in the previous example. Hence, in this case $a_t \in \mathbb{R}$. This so-called continuum bandit problem \citep{Bubeck2011b} has many practical applications. Here we provide an example of an alternative strategy for selecting the actions in such a setting. The term ``Lock-in Feedback" has been coined for this policy, which is described in detail in \citet{Kaptein2014d}. The basic idea of the policy is to oscillate the values of the actions at a known frequency and to amplify this frequency in the observed rewards. Next, the noise can be integrated out, which produces a result that---given mild assumptions regarding the function relating the reward and the action, which we denote $r = f(a)$---is directly proportional to the first derivative of $f()$. Subsequently, this first derivative can be applied, using a gradient-ascent-type algorithm, to move a step towards the maximum of $f()$.\footnote{Note, however, that Lock-in Feedback  does not attain asymptotically optimal performance due to its constant exploration.}

Lock-in Feedback is appealing because the experimenter does not need to specify $f()$ explicitly---as we did in the previous example---and the allocation policy has proved to be robust in cases of concept drift (e.g., a situation in which $f()$ changes over time). Lock-in Feedback can be implemented as follows:
\begin{itemize}
\item \code{getcontext}: For the sake of simplicity we consider a case without contextual information.
\item \code{getaction}: The implementation of the decision step Lock-in Feedback is relatively simple using the \code{lif} library:
\begin{Code}
theta = self.get_theta(all_float=False)
Lif = lif.LiF(theta, x0=3.0, a=.5, t=20, gamma=.02, omega=1.0, \
  lifversion=1)
suggestion = Lif.suggest()
self.action["x"] = suggestion["x"]
self.action["t"] = suggestion["t"]
self.action["x0"] = suggestion["x0"]
\end{Code}
where we refer to the \code{lif} documentation at \url{http://nth-iteration-labs.github.io/streamingbandit/} for details regarding the parameters of the \code{lif} method. 

\item \code{getreward}: Rewards can be simulated as follows:
\begin{Code}
x = self.action["a"]
self.reward["r"] = -1 * pow((x - 5), 2)
\end{Code}
where clearly the highest reward is obtained when $a = 5$.

\item \code{setreward}: Finally, the summary step can be implemented using
\begin{Code}
theta = self.get_theta(all_float=False)
Lif = lif.LiF(theta, x0=3.0, a=.5, t=20, gamma=.02, omega=1.0, \
  lifversion=1)
Lif.update(self.action["t"], self.action["x"], self.reward["r"], \
  self.action["x0"])
self.set_theta(Lif)
\end{Code}

\end{itemize}

Running a simulation with $n=1000$ and $\texttt{seed}=43123$ gives:
\begin{Code}
{
    "experiment": "2c070b0c17",
    "simulate": "success",
    "theta": {
        "x0": "4.9885573624026183",
        "t": "1000",
        "Yw": 
        "[[981.0, 5.32735516395185, -0.0364325832561265], 
        [982.0, 4.8625703653723935, 0.0023573471591685955], 
        [983.0, 4.512596828979011, 0.11280696009422152], 
        [984.0, 4.599313553861246, 0.06234374986452693], 
        [985.0, 5.0430128577187805, -0.00010219719086955388], 
        [986.0, 5.435840666435192, -0.0851018295535888], 
        [987.0, 5.416689112014516, -0.07446606113542824], 
        [988.0, 5.00321866842203, -1.599797646219837e-07], 
        [989.0, 4.575643921281427, 0.07422661156312545], 
        [990.0, 4.527110675559689, 0.10305907582127445], 
        [991.0, 4.902338977243727, 0.0008184672724298606], 
        [992.0, 5.356344978818639, -0.046745403226967665], 
        [993.0, 5.471845324305438, -0.10767081744630806], 
        [994.0, 5.142633922830094, -0.00314257165629084], 
        [995.0, 4.671518435948291, 0.034171390252964014], 
        [996.0, 4.491614635071014, 0.128372350733584], 
        [997.0, 4.7684753555595965, 0.011794462261213126], 
        [998.0, 5.247511963923335, -0.01586222111864476], 
        [999.0, 5.488454341053454, -0.11925205127092639], 
        [1000.0, 5.26974690054797, -0.020460304127878852]]"
    }
}
\end{Code}
Where \code{Yw} and \code{t} are internals used to execute the policy, and \code{x0} represents the current location of the search algorithm; initialized at $3.0$ it is, with a value of $4.988$ after $t=1000$ observations, indeed close to the actual maximum of $5$.

The \code{libs.lif} library has already been applied successfully in various settings, as described, for example, in a recent paper investigating the use of the LIF algorithm to optimize scenarios in behavioral economics \citep{kaptein2016tracking}, and in another paper in which LIF is applied to the optimization of the physical features of an avatar in multiple dimensions in response to a continuous stream of ratings, provided by the participants of the experiment \citep{kaptein2016investigation}. In both settings, LIF proved admirably capable of finding, and locking into, optima---despite the considerable noise often inherent in such human-choice-related studies. Hence, \pkg{StreamingBandit} was used successfully in these settings to allocate, in real-time, experimental treatments to subjects in a social-science study.

\subsection{Nesting of policies}
\label{sec:nesting}

A further interesting use of \pkg{StreamingBandit} relates to the ability to nest multiple policies; this allows the user to (e.g.,) use an $\epsilon$-greedy strategy to decide between the use of Lock-in Feedback and BTS, as presented above. Here we provide an example of this nesting of policies in which we assume that the user has instantiated two experiments, one implementing $\epsilon$-first as described in section \ref{sec:usage}, and one implementing $\epsilon$-first as described in section \ref{subsec:egreedy}. We can now setup a third experiment that allocates interactions to either of these two experiments by referring to their \code{exp_id}'s\footnote{Note that including a non-existent \code{exp\_id} leads to errors in running the code. \pkg{StreamingBandit} does not explicitly check for such errors inside the code of the user. We have implemented an implicit call that can be used to check if the experiment is valid by using \code{exp\_nested.is\_valid()}.}. This can be achieved as follows:

\begin{itemize}
\item \code{getcontext}: We do not consider a context in this example
\item \code{getaction}: Let us assume that we wish to uniformly randomly allocate half of our interactions to the $\epsilon$-first experiment, and half of our interactions to the $\epsilon$-greedy experiment. This can be done using:
\begin{Code}
id1 = "275fc0a66" # The exp_id of E-First
id2 = "18aec502c2" # The exp_id of E-Greedy

choice = np.random.binomial(1,0.5)

# Run the e-first experiment
if choice == 0:
    exp_nested = Experiment(exp_id = id1)
    self.action = exp_nested.run_action_code(context = {})
    # We return the experiment number for later use
    self.action["experiment"] = id1
    # We re-compute the propensity based on the probability of picking
    # the nested experiment
    self.action["propensity"] = self.action["propensity"] * 0.5
# or, run the e-greedy experiment
else: 
    exp_nested = Experiment(exp_id = id2)
    self.action = exp_nested.run_action_code(context = {})
    self.action["experiment"] = id2
    self.action["propensity"] = self.action["propensity"] * 0.5
\end{Code}

\item \code{getreward}: Rewards can be simulated using the code we also introduced in Section \ref{subsec:egreedy}.

\item \code{setreward}: The summary step for these nested experiments can be implemented using:
\begin{Code}
# Based on the exp_id we know which experiment to update
exp_id = self.action["experiment"]

exp_nested = Experiment(exp_id = exp_id)
exp_nested.run_reward_code(context = self.context, \
  action = self.action, reward = self.reward)
\end{Code}
Which simply, based on the supplied \code{exp_id}, updates the correct experiment.
\end{itemize}

Running a simulation with $n=2$ and $\texttt{seed}=13214$, and the output set to \texttt{verbose}, gives:
\begin{Code}
{
    "data": {
        "0": {
            "theta": {},
            "context": {},
            "reward": {
                "value": 4.439687566610595
            },
            "action": {
                "propensity": 0.45,
                "experiment": "18aec502c2",
                "treatment": "treatment"
            }
        },
        "1": {
            "theta": {},
            "context": {},
            "reward": {
                "value": 7.559583564021055
            },
            "action": {
                "propensity": 0.25,
                "experiment": "275fc0a66",
                "treatment": "treatment"
            }
        }
    },
    "experiment": "1c57b6d641",
    "simulate": "success"
}
\end{Code}
Which shows that in the first interaction $\epsilon$-greedy was selected, which subsequently selected the treatment arm, and in the second interaction $\epsilon$-first was selected. Obviously, this functionality can be greatly extended to use any sequential decision policy to decide between any other policy. This nesting makes \pkg{StreamingBandit} a versatile tool; we illustrate a practical application of the nesting in Section \ref{sec:offpol}.

\subsection{Parallel evaluation of multiple policies}
\label{sec:offpol}

Whereas the nesting discussed in the previous section allows one to allocate different interactions to different policies, the example we provide here allows one to evaluate, using a measure of average reward for example, multiple bandit policies in parallel. The idea behind the parallel evaluation derives from recent work on the offline evaluation of bandit policies. \citet{Li2010a} show that one can evaluate multiple bandit policies offline by simply running through an existing data set of actions and rewards obtained using uniform random selections of the actions. For each interaction $t$ in the offline data set one uses a bandit policy to generate a proposal action $a'_t$, and if the randomly selected action at that point in time matches the proposal (thus $a'_t = a_t$), then the reward is used to update the estimated performance of the policy. If not, then the time point is discarded. This leads to an evaluation of the policy with an expected number of observations of $\frac{1}{k} T$, where $k$ is the number of possible actions and $T$ the total number of observations in the offline data set. Multiple offline evaluation runs can subsequently be used to estimate and compare the expected performance of different policies.

Here we extend this idea to the parallel evaluation of multiple bandit policies. The implementation in \pkg{StreamingBandit} to compare, in parallel, the performance of the $\epsilon$-first and $\epsilon$-greedy experiments as introduced above is surprisingly straightforward:
\begin{itemize}
\item \code{getcontext}: For simplicity we again consider an empty context.
\item \code{getaction}: In the decision step an action is chosen at random:
\begin{Code}
self.action["treatment"] = random.choice(["control","treatment"])
\end{Code}

\item \code{getreward}: Rewards can again be simulated using the code we also introduced in Section \ref{subsec:egreedy}.
\item \code{setreward}: Finally, after generating a reward, the summary step for the parallel evaluation of the policies is given below, where we again insert comments in the code to improve readability:
\begin{Code}
# Create a list of experiments / policies to evaluate
policies = ["18aec502c2", # E-Greedy
                 "275fc0a66"] # E-First

# For each experiment
for exp_id in policies:

    # Initialize the experiment:
    exp_nested = Experiment(exp_id)
    
    # Compute the suggested action:
    suggestion = exp_nested.run_action_code(context = {})
    
    # See if the suggested action matches the actual action:
    if suggestion["treatment"] == self.action["treatment"]:
        
        # And if so store the performance of the policy:
        mean = base.Mean(self.get_theta(key = "policy_means", 
        				value = exp_id))
        mean.update(self.reward["value"])
        self.set_theta(mean, key = "policy_means", value = exp_id)
        
        # And finally update the policy:
        exp_nested.run_reward_code(context = {}, 
        		action = self.action, reward = self.reward)
\end{Code}
This code implements Algorithm $2$ in \citep{Li2010a}.
\end{itemize}

Running a simulation with $n=250$ and $\texttt{seed}=43123$ using the above specification gives:
\begin{Code}
{
    "theta": {
        "policy_means:275fc0a66": {
            "m": "5.243151928222057",
            "n": "114"
        },
        "policy_means:18aec502c2": {
            "m": "6.050783848360902",
            "n": "114"
        }
    },
    "experiment": "270ed59474",
    "simulate": "success"
}
\end{Code}
This output shows that, in this test run, the average reward of the $\epsilon$-greedy policy is slightly higher than that of the $\epsilon$-first policy. This is due to the fact that $\epsilon$-first has a random exploration phase of $n = 100$. Since both policies now only have had $114$ accepted actions, $\epsilon$-first will have explored much more than $\epsilon$-greedy and will choose the suboptimal action more, resulting in a lower average reward.


\section{Applied usage}
\label{sec:practice}

In this Section, we describe some of the practical applications of \pkg{StreamingBandit}. First, we explore its use in assessing the effects of discounts in online selling; this small, initial trial highlights the simple use of \pkg{StreamingBandit} to collect data in-the-field. Second, we introduce its use in a social-science experiment.

\subsection{Online marketing}

\pkg{StreamingBandit} was used by an online cash-refund company to examine the effects of their pricing scheme. The company offers customers the opportunity to sign up for a refund program. After signing up they are provided with discounts, in the form of a cash refund, as long as their online purchases are carried out through the online platform. The refund company has negotiated different agreements with a large number of different e-commerce stores, and the discount percentages they have obtained vary from store to store. By default, the refund company offers half of its negotiated discount to the customer, and takes the other half as a fee for its services. However, it has no clear idea as to whether this 50/50 (or $\frac{1}{2}$) split is optimal in the sense that it maximizes its profit, which is influenced by the total number of purchases, the size of the purchases, and the way in which the negotiated discount is split between the company and the customer.

The company set up \pkg{StreamingBandit} to explore the effects of the different splits---in their definition running from $0$ to $1$ where $1$ means that the total negotiated discount is fully passed on to the customer and $0$ means that all of it is retained by the company---on their resulting profits. Here we present a simple implementation of the random exploration of different splits that the company carried out for a very small number of $n=103$ unique customers in one specific store. The implementation was as follows:

\begin{itemize}
\item \code{getcontext}: Because this is a field exploration, the context was provided by the participating company. It consisted of a JSON object containing the \code{maxpercentage}, which contained the negotiated discount for the specific store that was viewed by a customer. It looked like this:
\begin{Code}
{"context" : {"maxpercentage" : 8.5}}
\end{Code}
where the \code{maxpercentage} for the specific store from which our presented data originated was always $8.5\%$. However, our implementation described below is able to address changing maximum percentage(s) between different stores. Note that this can be simulated in \pkg{StreamingBandit} using the following \code{getcontext} code:
\begin{Code}
self.context["maxpercentage"] = numpy.random.uniform(1,10)
\end{Code}

\item \code{getaction}: The implementation of the decision step was straightforward since the company initially set out merely to examine the effects of random fluctuations of the discounts offered. The implementation was as follows:
\begin{Code}
maxpercentage = self.context['maxpercentage']
split = np.random.uniform()
discount = split * maxpercentage
self.action['split'] = split 
self.action['discount'] = discount
\end{Code}
Here, first the \code{maxpercentage} is retrieved. Next, a split is computed ($\texttt{split} \sim \texttt{unif}(0,1)$), after which the percentage discount to be offered to the customer is computed and then both the \code{split} and the actual \code{discount} are returned in the \code{action} object.

\item \code{getreward}: The online platform would display the computed \code{discount} to the visiting customer, and subsequently a reward would be generated by virtue of the customer's purchasing one or multiple products resulting in a \code{revenue}. The online platform returns both the \code{revenue} as well as the \code{split} and \code{discount}. 
This can be simulated using:
\begin{Code}
self.action['split'] = self.action['split']
self.action['discount'] = self.action['discount']
self.reward['revenue'] = numpy.random.uniform(0,100)
\end{Code}

\item \code{setreward}: Finally, given that the aim of the company was merely to collect data on the effect of the changing splits, it did not need any \code{setreward} code because \pkg{StreamingBandit} automatically logs all the data that is received with a \code{setreward} call.
\end{itemize}

This simple implementation allowed the refund company to vary the split randomly (instead of using the current de-facto $\frac{1}{2}$ split) and to log the resulting revenue. 

\begin{figure}[h!]
  \centering
    \includegraphics[width=.75\textwidth]{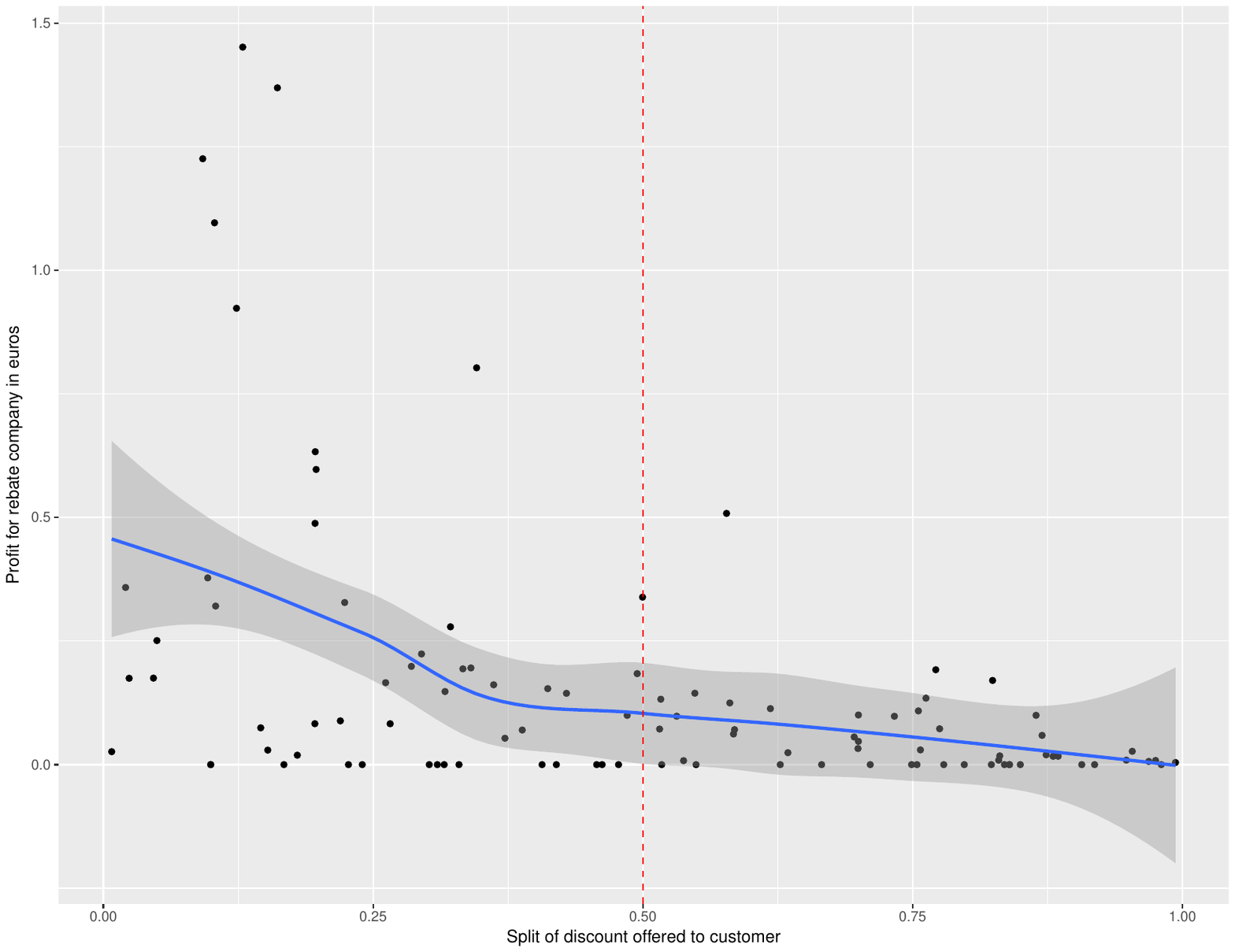}
      \caption{Overview of the effect of the offered split of the discount on the profit of the refund company in euros. Data collected using random selection of the refund percentage using \pkg{StreamingBandit}. The Figure presents data on $n=103$ unique customers. The dashed red line represents the company's current $\frac{1}{2}$ split.}
       \label{fig:BB}
\end{figure}

Figure \ref{fig:BB} provides an overview of the relation between the suggested \code{split} and the resulting profit in euros of the refund company. The \code{profit} for the rebate-company is defined as the maximum discount percentage ($8.5\%$) times one minus the split (between $0$ and $1$), times the revenue. Each dot represents one completed purchase by one customer (possibly containing multiple products). Note that we while limit the presented results here to a single e-commerce store, the store sells multiple products and hence the revenue per customer can vary greatly. It seems from the limited data of these $n=103$ unique customers for a single store that a high customer-refund offer---but as a result a low margin for the company---leads to low profits, whereas an offer that is significantly below the current $\frac{1}{2}$ split increases the company's profits.

The company intends to use \pkg{StreamingBandit}, now that the software is integrated into its current online service, to experiment with different sequential allocation schemes that offer different splits between competing stores or between different customers. Using the random data and an adaption of the offline evaluation method developed by \citep{Li2010a} (also described in section \ref{sec:offpol}), the company hopes to find the policy that has the best model fit on their data. Note that here every step towards solving this statistical decision problem involves using \pkg{StreamingBandit}---from gathering data, to policy evaluation, to the final, live setting. This provides a simple example of the utility of \pkg{StreamingBandit} for field trials of bandit policies.

\subsection{Social science experiment}

The second applied use of \pkg{StreamingBandit} we present concerns a social-science experiment examining the decoy effect \citep[see][for a full description of the experiment]{kaptein2016tracking}. In short, the decoy effect states that people may be persuaded to switch from one offer to another by the presence of a third option (the decoy) that, rationally, should have no influence on the decision-making process. For example, when asked to choose between a laptop with a good battery but a poor memory and a laptop with a poor battery but a good memory, people seem to shift their preference between the two if the offer is accompanied by a third laptop, the decoy, that has a battery as good as the latter but an even worse memory, and hence should in any case be an irrelevant option. The placement of the decoy in the product-attribute space is heavily studied in the literature: researchers manipulate the exact battery life in hours and the RAM in GB of the decoy laptop, and study the resulting choices that people make.

\citet{kaptein2016tracking} used \pkg{StreamingBandit} to study whether Lock-in Feedback, the sequential optimization scheme introduced in section \ref{sec:lif}, can be used to find the optimal placement of the decoy--- only considering changes on one dimension. The authors considered not only the laptop scenario but also eight different decoy scenarios. The study was carried out online using a \code{drupal}-based survey, which communicated with \pkg{StreamingBandit} to implement the allocation of the exact positioning of the decoy. The researchers allocated participants to one of $3$ between-subject conditions using \pkg{StreamingBandit}:
\begin{enumerate}
\item Baseline: participants in this condition were presented with a binary choice between two products, and no decoy was present. This was implemented by sending an \code{action} with	\code{\{"decoy":"none"\}} response to the survey front-end.
\item Random: participants in this condition were presented with a random positioning of the decoy. The range of possible values of the random positioning depended on the specific scenario, and were hard- coded and retrieved using the \code{scenario} supplied in the context.
\item Lock-in Feedback: participants in this condition were presented with a value of the decoy that depended on the previous interactions of other participants. The Lock-in Feedback algorithm was used to suggest a new placement each time a participant viewed a product. Subsequently, the (binary) choice made by the participant was used to update the algorithm in the \code{setreward} stage. We refer the reader to \citep{kaptein2016tracking} for details and for the exact settings of the tuning parameters.
\end{enumerate}

\begin{figure}[h!]
  \centering
    \includegraphics[width=.75\textwidth]{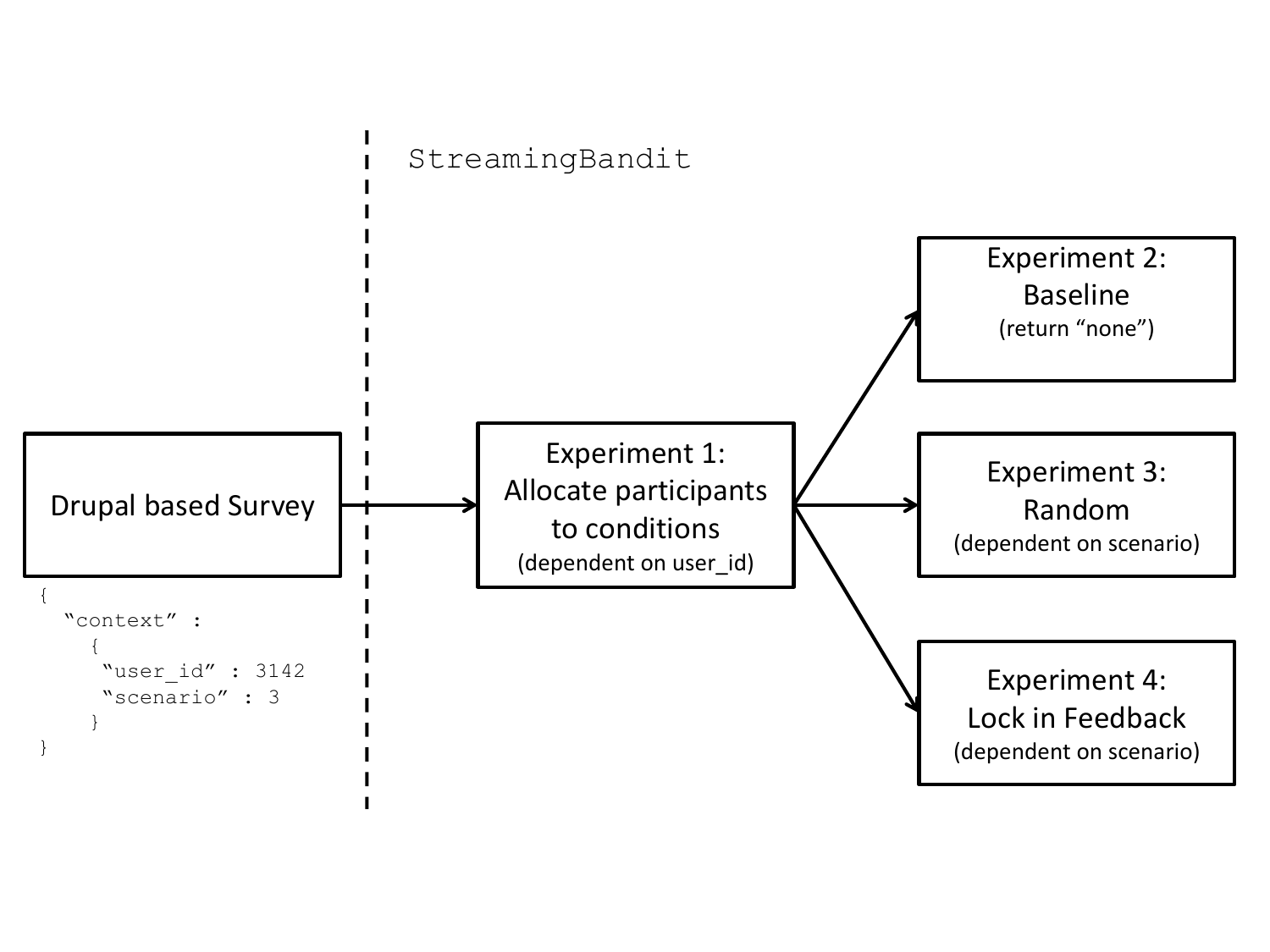}
      \caption{Schematic setup of the $4$ \pkg{StreamingBandit} experiments used to realize the data-collection in \citep{kaptein2016tracking}.}
      \label{fig:lif}
\end{figure}

Figure \ref{fig:lif} presents an overview of the setup of this study. A number of the details of the implementation are covered in earlier sections of this paper: the implementation of both the baseline and the random condition are straightforward, with \code{self.action["decoy"] = "none"} and \code{self.action["decoy"] = np.random.uniform(low,high)}, respectively, as the core \code{getaction} implementations. In the latter implementation the \code{low} and \code{high} bounds were implemented as a simple list indexed by the \code{scenario} number. Finally, the Lock-in Feedback condition was implemented as presented in Section \ref{sec:lif}, the only exception being that the \code{theta} was stored independently for each \code{scenario}. Hence, the novel part of the implementation of this study is the persistent allocation of participants to one of the three conditions; this was achieved in experiment 1 in Figure \ref{fig:lif} by using the following
\code{getaction} code:

 \begin{Code}
if not("condition" in self.get_theta("user_id", self.context["user_id"])):
    self.action["note"] = "First allocation"
    draw = random.choice(["baseline", "random", "lockin"])
    self.set_theta({"condition":draw}, "userid", self.context["userid"])

self.action["condition"] = 
    self.get_theta("userid", self.context["userid"])["condition"]
\end{Code}
which assigns participants randomly to one of the three conditions persistently based on the \code{user_id} supplied in the context to the \code{getaction} call.\footnote{Note that the actual implementation in the study differed slightly to allow for unequal sample sizes in each of the conditions. In addition, the baseline and random conditions where manually removed after sufficient data had been collected.}

The data resulting from this experiment are available at \url{http://dx.doi.org/10.7910/DVN/FCHU0J}. This field implementation provides a prime example of the use of \pkg{StreamingBandit} both for the allocation of participants to conditions in (web based-) experiments, as well as in sequential decision policies such as Lock-in Feedback in such experiments.


\section{Conclusion and Future work}
\label{sec:discussion}
This paper presented \pkg{StreamingBandit} a RESTful web application that enables researchers to develop, evaluate, and deploy CMAB policies in online experiments and field studies. By making \pkg{StreamingBandit} publicly available we hope to contribute to the more extensive use of such policies to solve statistical decision problems. The software could help in extending the currently prevailing use of basic random assignment to the use of more refined strategies throughout the social and medical sciences. To that effect, we started out with a clarification of the design rationale behind \pkg{StreamingBandit}. We explained	our decision to split up the summary and the decision step of a policy---a split meant to encourage the implementation of computationally efficient online policies. We subsequently illustrated \pkg{StreamingBandit}'s versatility and flexibility in a number of examples, and we concluded with two case studies in which we used \pkg{StreamingBandit} to run field experiments.

We are currently aware of a number of limitations of \pkg{StreamingBandit}. First, as of now, \pkg{StreamingBandit} still runs single-threaded. Although parallelization for larger-scale applications ought to be relatively easy to implement on the level of policies, it may prove substantially harder within policies. Nevertheless, by forcing policies online by design, and using state-of-the-art web technology for its back end, \pkg{StreamingBandit} is already more than capable of being deployed in a multitude of small-to-medium-sized field trials. We are of the opinion that parallelization is an obvious next step in \pkg{StreamingBandit}'s development, ensuring its future scalability.

Second, in some applications we find that certain types of reward manifest themselves faster than others. In one instance of the use of \pkg{StreamingBandit}, for example, the decision to reject a loan to a customer after an application had been submitted to the firm was much faster than the decision (and subsequent confirmation) to accept the customer. Such an asymmetric delay might bias learning and thus needs to be addressed. Currently, we do not provide an off-the-shelf solution to this problem---admittedly because it is thus far unclear to us how to address the problem in general---hence users will need to resort to custom implementations of the \code{getaction} and \code{setreward} codes to deal with this issue.

Finally, our current CMAB libraries and toolkit still offer ample room for improvement and extension. Outside of the currently implemented methods, there are many more policies that address the exploration-exploitation trade-off in various settings. In that respect, we hope and expect the open-source nature of \pkg{StreamingBandit} to be conducive to the continued growth of the platform, encouraging researchers to implement, test, and disseminate new and existing bandit policies and algorithms.


\bibliography{library}
\clearpage

\section*{Appendix A: Setting up an experiment}
\label{app:ui}

\renewcommand{\thefigure}{A\arabic{figure}}

\setcounter{figure}{0}

This Appendix introduces the front-end of \pkg{StreamingBandit}.\footnote{Which can be found at \url{https:// github.com/Nth-iteration-labs/streamingbandit-ui}.} We will how show to get from the login screen to setting up your first simulation using one of the default experiments.

\begin{figure}[H]
  \centering
  \includegraphics[width=.75\textwidth]{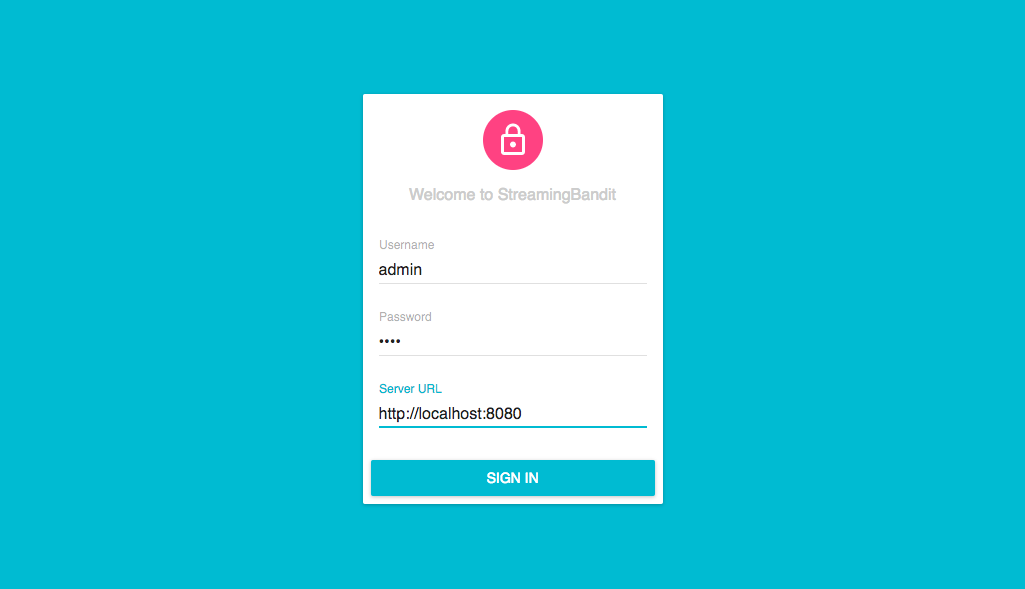}
  \caption{The login screen. Here you can set the URL for the back-end and login.}
  \label{A1}
\end{figure}

First, when you have set up the front-end (using e.g., the available \pkg{Docker} container), go to the login screen in your browser (for the \pkg{Docker} container this would be \url{http://localhost} or the Docker-set IP address) as shown in Figure \ref{A1}.

\begin{figure}[H]
  \centering
    \includegraphics[width=.75\textwidth]{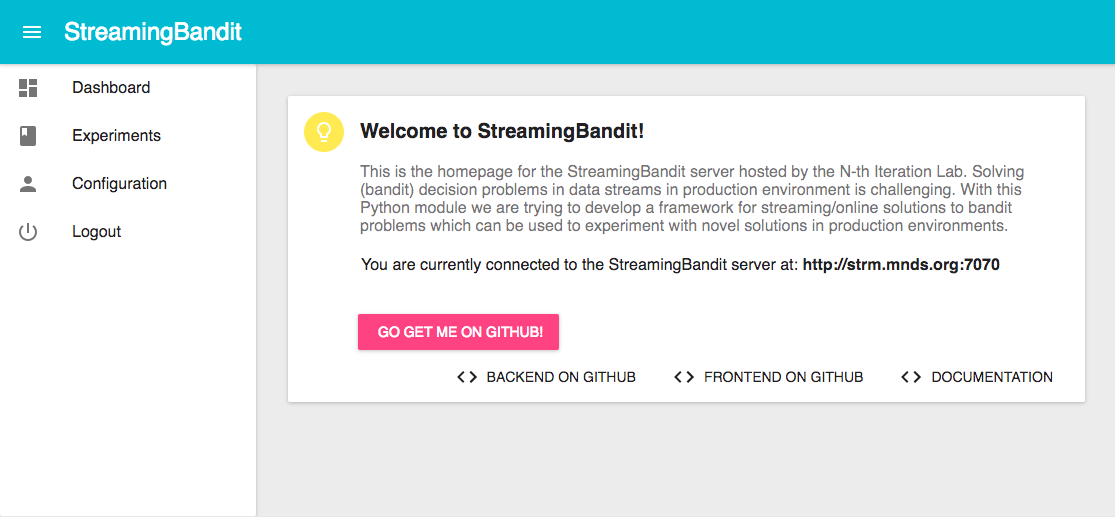}
    \caption{The dashboard of \pkg{StreamingBandit}, here you can navigate to the list of experiments and find some extra information.}
    \label{A2}
\end{figure}

\begin{figure}[H]
  \centering
    \includegraphics[width=.75\textwidth]{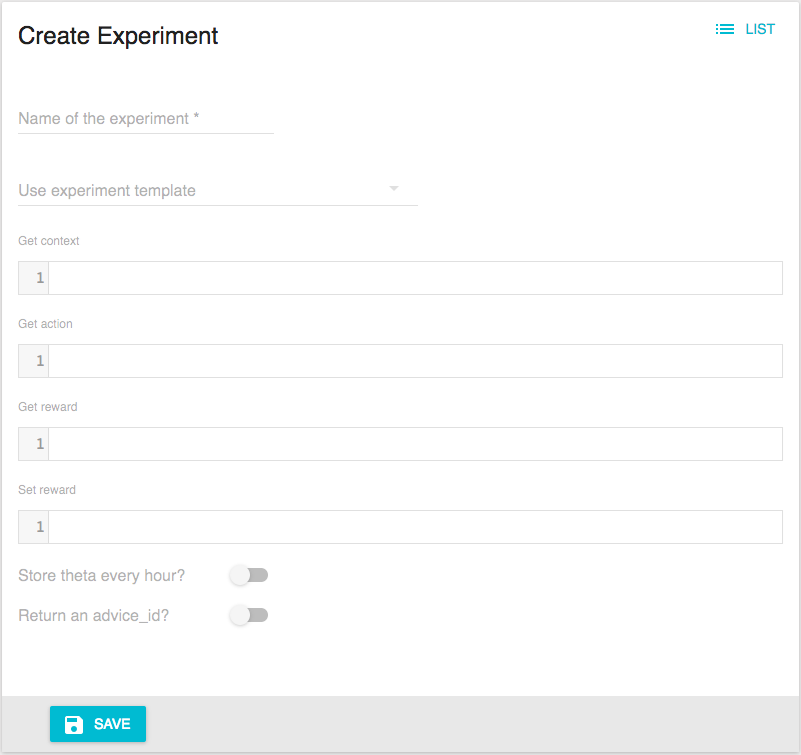}
     \caption{This is an empty form for an experiment, which is normally shown inside the dashboard. Here you can type a name, choose a default and edit the code for the experiment.}
    \label{A4}
\end{figure}

After logging in, you will find the dashboard as in Figure \ref{A2}. To show all the active experiments, click on \code{Experiments}. This will bring you to an environment as shown in Figure \ref{A4}. Continue clicking on the \code{Create} button, which will give you an empty \code{Create Experiment} field, as in Figure \ref{A4}.

\begin{figure}[H]
  \centering
    \includegraphics[width=.75\textwidth]{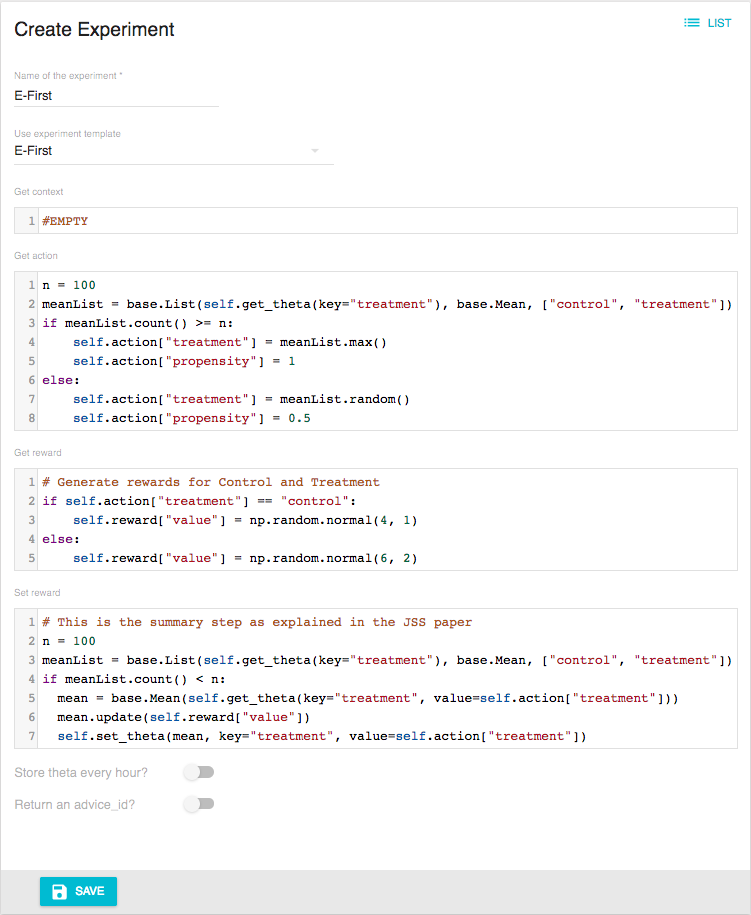}
     \caption{We have selected the E-First template, which is automatically filled in the correct fields. Pressing the \code{Save} button will create the experiment.}
    \label{A5}
\end{figure}

On the creation page you can fill in a name, for example \code{E-First}, and select a default experiment from the \code{Use experiment template} list. Selecting the default $\epsilon$-first experiment, will end up with a filled-in form, as in Figure \ref{A5}. Next, clicking on the \code{Save} button will save the experiment in the database.

\begin{figure}[H]
  \centering
    \includegraphics[width=.75\textwidth]{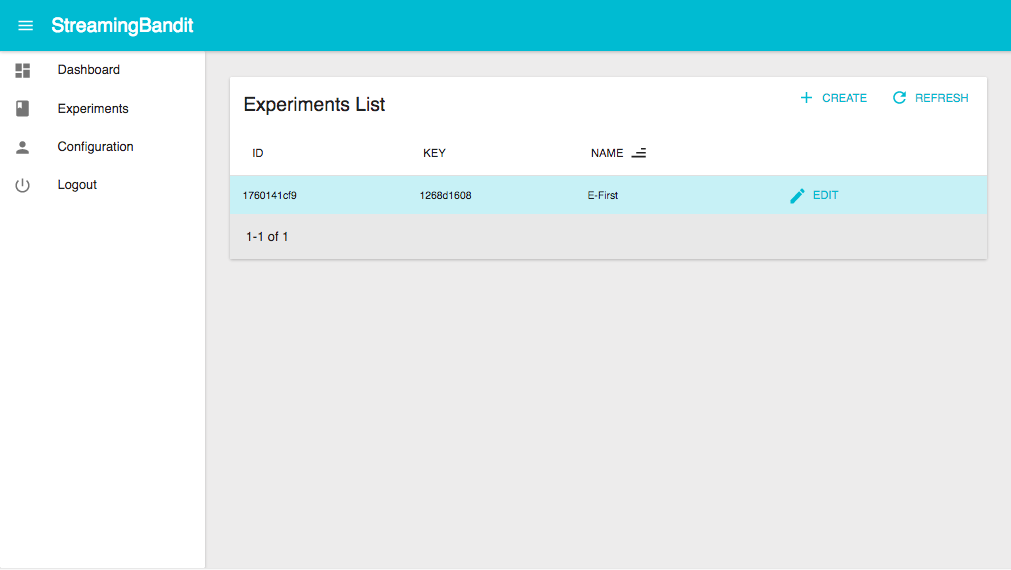}
     \caption{If you return to the dashboard you can view and edit your newly created experiment (and the associated ID and key).}
    \label{A6}
\end{figure}

\begin{figure}[H]
  \centering
    \includegraphics[width=.75\textwidth]{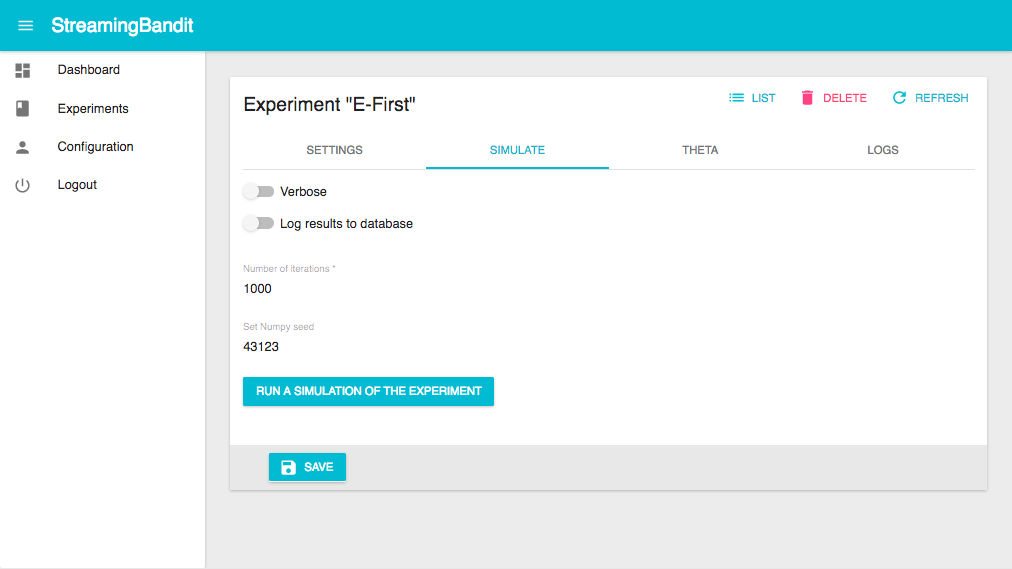}
     \caption{Here you can see the simulation panel, which you can use to easily run a simulation of the experiment. You can set the speed and the the number of iterations, log the results to the database and even show verbose results. Click \code{Run a simulation of the experiment} to run a simulation and get an output of the results.}
    \label{A7}
\end{figure}

\begin{figure}[H]
  \centering
    \includegraphics[width=.75\textwidth]{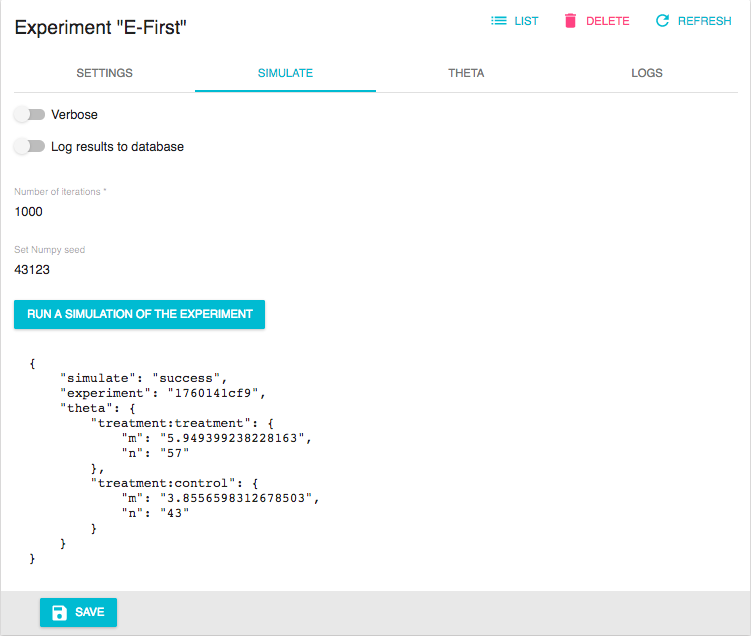}
     \caption{Run a simulation and look at the output of the results.}
    \label{A8}
\end{figure}

\begin{figure}[H]
  \centering
    \includegraphics[width=.75\textwidth]{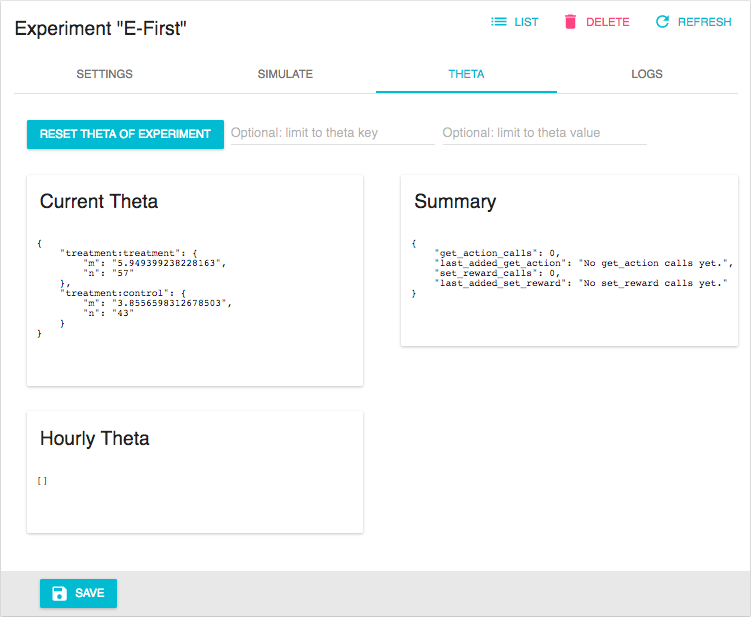}
     \caption{The theta panel shows the current state of $\theta$ and other information.}
    \label{A9}
\end{figure}

When the experiment has been created the dashboard (Figure \ref{A6}) shows that the experiment is active and has an ID and key assigned. Clicking on the \code{Edit} button will take you back to the settings of the experiment. Now you can choose to go to the \code{Simulate} tab as displayed in Figure \ref{A7}. After filling in $1000$ for the number of iterations and $43123$ as the seed you can click \code{Run a simulation of the experiment}, which will give a result as in Figure \ref{A8}. Finally, you can click on the \code{Theta} tab and inspect the parameters that are stored in the database (Figure \ref{A9}). Here you can also download the data that has been logged for the current experiment.

\begin{figure}[H]
  \centering
    \includegraphics[width=.75\textwidth]{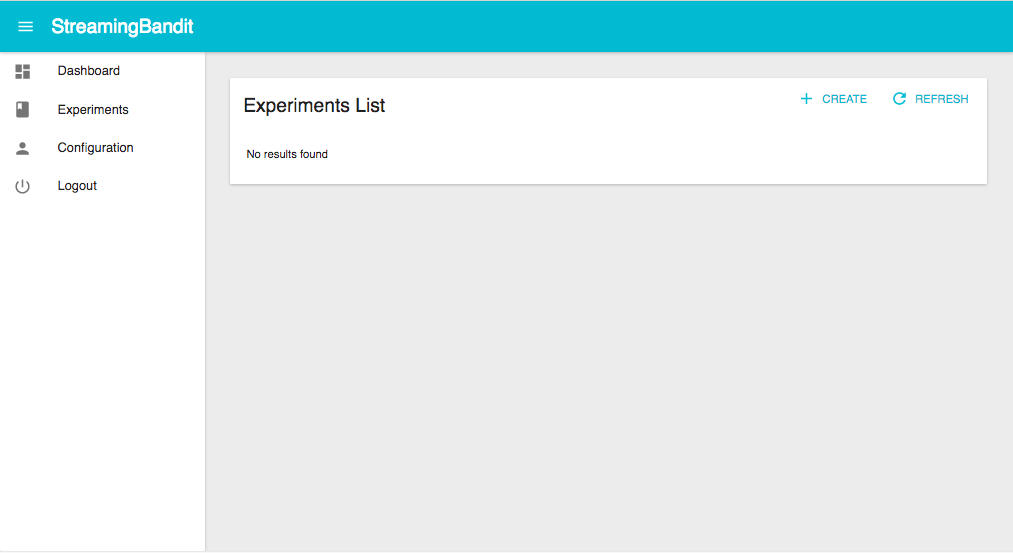}
     \caption{This screenshot shows an empty list of experiments. You can start creating an experiment by clicking on the \code{Create} button.}
    \label{A3}
\end{figure}

\end{document}